\documentclass[12pt,a4paper]{extarticle}

\pdfoutput=1

\usepackage{amsmath}
\usepackage{amssymb}
\usepackage{amsthm}

\usepackage{graphicx}

\usepackage[utf8]{inputenc}
\usepackage[english]{babel}

\makeatletter
\numberwithin{equation}{section}
\numberwithin{figure}{section}
\newtheorem{thm}{\protect\theoremname}[section]
\newtheorem{prop}[thm]{\protect\propositionname}

\newtheorem{remark}[thm]{\protect\remarkname}
\newtheorem{example}[thm]{\protect\examplename}
\makeatother

\providecommand{\propositionname}{Proposition}
\providecommand{\corollaryname}{Corollary}
\providecommand{\theoremname}{Theorem}
\providecommand{\remarkname}{Remark}
\providecommand{\examplename}{Example}

\newcommand*{\ol}{\overline}
\newcommand*{\PP}{\mathsf{P}}
\newcommand*{\EE}{\mathsf{E}}

\newcommand*{\Var}{\mathsf{Var}}

\newcommand*{\cF}{\mathcal{F}}
\newcommand*{\cG}{\mathcal{G}}
\newcommand*{\arginf}{\operatornamewithlimits{arg\,min}}

\newcommand*{\sgn}{\operatorname{sgn}}
\newcommand*{\maxj}{\max_{0\le j\le J}}
\newcommand*{\const}{\mathrm{const}}

\newcommand*{\NN}{\mathbb{N}}

\begin{document}
\title{Regression-based complexity reduction of the nested Monte Carlo methods \thanks{Stefan H\"afner thanks the Faculty of Mathematics of the University of Duisburg-Essen, where this work was carried out.}}
\author{Denis Belomestny\thanks{University of Duisburg-Essen, Essen, Germany; E-mail address: \texttt{denis.belomestny@uni-due.de}}\and Stefan H\"afner\thanks{PricewaterhouseCoopers GmbH, D\"usseldorf, Germany; E-mail address: \texttt{stefan.haefner@pwc.com}}\and Mikhail Urusov\thanks{University of Duisburg-Essen, Essen, Germany; E-mail address: \texttt{mikhail.urusov@uni-due.de}}}
\date{\phantom{March 10, 2018}}
\maketitle

\begin{abstract}
In this paper we propose a novel dual regression-based approach for pricing American options. This approach reduces the complexity of the nested Monte Carlo method  and has especially simple form for  time discretized  diffusion processes. We analyse the complexity of the proposed approach both in the case of fixed and increasing number of exercise dates. The method is illustrated by several numerical examples.


\end{abstract}

\section{Introduction}
\label{berm:sec:intro}
In contrast to European options, which may be exercised only at a fixed date, an American option grants its holder the right to select the time
at which to exercise the option. A general class
of American option pricing problems can be formulated through an \ensuremath{\mathbb{R}^{d}}-valued
$(\mathcal F_t)$-Markov process \ensuremath{(X_{t})_{t\in[0,T]}}
with a deterministic starting point $X_0=x_0\in\mathbb{R}^d$
defined on a filtered probability space \ensuremath{(\Omega,\mathcal{F},(\mathcal{F}_{t})_{0\leq t\leq T},\PP).}
Let us recall that each \ensuremath{\mathcal{F}_{t}}
 is a \ensuremath{\sigma}-algebra
 of subsets of \ensuremath{\Omega},
and \ensuremath{\mathcal{F}_{s}\subseteq\mathcal{F}_{t}\subseteq\mathcal{F}}
 for \ensuremath{s\leq t.} We first consider options admitting a finite set of exercise
opportunities \ensuremath{0=t_{0}<t_{1}<t_{2}<\ldots<t_{J}=T,} called Bermudan options, with corresponding Markov chain 
\[
X_{j}:=X_{t_{j}},\quad j=0,\ldots, J.
\]
This option pays \ensuremath{g_{j}(X_{j})}, if exercised at time \ensuremath{t_{j},\, j=0,\ldots,J}, for some known Borel-measurable functions \ensuremath{g_{0},\ldots,g_{J}}
mapping \ensuremath{\mathbb{R}^{d}}
into \ensuremath{[0,\infty)}.
Below we assume that $g_j(X_j)\in L^2$ for all~$j$.
Let \ensuremath{\mathcal{T}_{j}}
denote the set of stopping times taking values in \ensuremath{\{j,j+1,\ldots,J\}}. As a standard result in the theory of contingent claims, the equilibrium price \ensuremath{v^*_{j}(x)}
 of the Bermudan option at time \ensuremath{t_{j}}
 in state \ensuremath{x},
 given that the option was not exercised prior to \ensuremath{t_{j}},
 is its value under the optimal exercise policy
\begin{eqnarray*}
v_{j}^{*}(x)=\sup_{\tau\in\mathcal{T}_{j}}\EE[g_{\tau}(X_{\tau})|X_{j}=x],\quad x\in\mathbb{R}^{d}.
\end{eqnarray*}
Clearly, any given stopping
rules \(\tau_j\in\mathcal{T}_{j}\) are
generally suboptimal and give us lower bounds
\begin{eqnarray*}
v_{j}(x):=\EE[g_{\tau_j}(X_{\tau_j})|X_{j}=x]\leq v_{j}^{*}(x),
\quad j=0,\ldots,J,
\end{eqnarray*}
for the option price.

By now, there are well established algorithms
that produce tight lower bounds
for prices of Bermudan options.
For instance, the computationally efficient
Longstaff and Schwartz \cite{LS} algorithm
based on nonparametric regression is widely used.
It is crucial especially in high-dimensional problems,
where explicit formulas for option prices are typically unavailable
even in the simplest Black-Scholes framework.
Moreover, in addition to good lower bounds
it is important to have tight upper bounds for the prices
because only in the case when we have both bounds,
we have reliable confidence intervals for the
unknown true price as well as we know the magnitude of the error
in our approximations.
Even when an estimate of the variance of
a lower bound for the option price is available,
we usually cannot construct a reliable
confidence interval for the unknown true option price
because we typically do not know the magnitude of the bias.
An upper bound for the true price could be generated
from any given exercise policy using the following dual approach,
which was proposed in Rogers \cite{Ro} and Haugh and Kogan \cite{HK}.
For any $0\leq i\leq J$ and any
supermartingale $(Y_{j})_{i\leq j\leq J}$ with $Y_{i}=0$, it holds 
\begin{align}
v^*_{i}(X_{i})=\sup_{\tau \in \mathcal{T}_{i}}\EE\left[ g_{\tau
}(X_{\tau })|\mathcal{F}_{i}\right]
&
\leq \sup_{\tau \in \mathcal{T}_{i}}\EE\left[ g_{\tau }(X_{\tau })-Y_{\tau }|\mathcal{F}_{i}\right]
\notag\\
&
\leq \EE\left[ \max_{i\leq j\leq J}\left(
g_{j}(X_{j})-Y_{j}\right) |\mathcal{F}_{i}\right]
\label{berm:dual}
\end{align}
(we now use the shorthand $\cF_j:=\cF_{t_j}$).
Therefore the right-hand side of~\eqref{berm:dual} provides an upper bound for $v^*_{i}(X_{i})$. It can
be derived that both inequalities in~\eqref{berm:dual}
are equalities for the martingale
part of the Doob-Meyer decomposition of the price process
$(v^*_j(X_j))_{i\le j\le J}$
\begin{equation*}
Y^*_{i}=0,\quad Y^*_{j}=\sum_{l=i+1}^{j}\left( v^*_{l}(X_{l})-\EE\left[ v^*_{l}(X_{l})|
\mathcal{F}_{l-1}\right] \right),
\quad j=i+1,\ldots,J.
\end{equation*}
In fact, \(Y^*\) satisfies the following even stronger almost sure identity
\begin{eqnarray}
\label{berm:eq:as_doob}
v^*_{i}(X_{i})=\max_{i\leq j\leq J}\left(
g_{j}(X_{j})-Y^*_{j}\right),\quad a.s.
\end{eqnarray}
(see \cite{SchZhH2013}). 
The duality representation provides a simple way to estimate the Snell
envelope from above, using approximations  $(v_{i}(X_i))$ for the value functions \((v^*_{i}(X_i)).\)
Let $Y$ be a martingale defined via
\begin{equation}
Y_{0}=0,\quad
Y_{j}=\sum_{l=1}^{j}
\left(v_{l}(X_{l})-\EE\left[ v_{l}(X_{l})|\mathcal{F}_{l-1}\right]\right),
\quad j=1,\ldots,J.
\label{berm:eq:07092016a1}
\end{equation}
Then, for $i=0$, we get that
\begin{align}
\label{berm:eq:repr_upper}
V_0:=v_{0}(x_0)
=\EE\left[ \max_{0\leq j\leq J}\left(
g_{j}(X_{j})-Y_j \right) \right]
\end{align}
is an upper bound for \(V^*_0:=v^*_0(x_0)\).

Another approach to construct upper bounds is based
on the so-called discrete time early exercise premium representation
$$
v^*_{0}(x_0)=\EE\left[ g_J(X_{J}) +\sum_{l=1}^{J}
(g_{l-1}(X_{l-1})-\EE[ v^*_{l}(X_{l})|\mathcal{F}_{l-1}])^{+}\right],
$$
which was first established in~\cite{belomestny2006monte}.
Then, for any lower approximation \(v_l(X_l)\) with \(v_l(X_l)\leq v^*_l(X_l),\) \(l=1,\ldots, J,\) a.s., we get an upper bound
\begin{equation}\label{eq:upper_eep}
U_{0}=\EE\left[ g_J(X_{J}) +\sum_{l=1}^{J}
(g_{l-1}(X_{l-1})-\EE[ v_{l}(X_{l})|\mathcal{F}_{l-1}])^{+}\right],
\end{equation}
i.e.\ $U_0\ge V_0^*$.
There are examples, where the upper bound $U_0$
is more accurate than the dual upper bound $V_0$,
and there are opposite examples
(see Section~2.4 in~\cite{BMS:2009}).

In this paper, we suggest a novel nonparametric regression algorithm to construct computationally efficient approximations for the conditional expectations involved
in~\eqref{berm:eq:07092016a1}
and~\eqref{eq:upper_eep}.
Nonparametric regression algorithms like that of Longstaff and Schwartz have become among the most successful and widely used methods for approximating the values of American-style (Bermudan) options, in particular for high-dimensional problems. Due to their popularity, the analysis of the convergence properties of these types of Monte Carlo algorithms is a problem of fundamental importance in applied probability and mathematical finance,
see e.g.\ Cl{\'e}ment, Lamberton and  Protter \cite{CLP2002}, Zanger \cite{zanger2013quantitative} and references therein.
Here we rigorously analyse the convergence properties of the proposed regression algorithm and derive its complexity.
To this end, we first establish in Section~\ref{berm:sec:ns}
a new $L^2$ error bound for the nested simulations approach
based either on~\eqref{berm:eq:repr_upper}
or on~\eqref{eq:upper_eep},
which turns out to be instrumental both
for understanding how to improve the standard nested estimators
and for the error analysis of the proposed algorithm
(the latter is constructed and studied in detail in later sections).
The performance of our algorithm is illustrated
by the example of max-call Bermudan options.

\section{Nested simulations approach}
\label{berm:sec:ns}
The nested simulations approach  for computing \(V_0\)
of~\eqref{berm:eq:repr_upper}
relies on the approximation of the conditional expectations
in~\eqref{berm:eq:07092016a1} via (nested) Monte Carlo.
This approach was first proposed in Andersen and Broadie \cite{AB} for the computation of the dual upper bound~\eqref{berm:eq:repr_upper}. Let us describe this method in more detail. Fix some natural numbers \(N_d\) and~\(N\).
The dual nested simulations approach consists in using the estimate
$$
V_{N,N_d}=\frac{1}{N}\sum_{n=1}^N\left[ \max_{0\leq j\leq J}\left(
g_{j}(X^{(n)}_{j})-Y_{j,n,N_d} \right) \right],
$$
where 
\begin{eqnarray*}
Y_{j,n,N_d}=\sum_{l=1}^{j}\left(v_{l}(X^{(n)}_{l})-\frac{1}{N_d}\sum_{n_d=1}^{N_d} v_{l}(X^{(n_d,n)}_{l}) \right),
\quad j=0,\ldots,J,
\end{eqnarray*}
($\sum_1^0 := 0$),
\((X_l^{(1)},\ldots,X_l^{(N)})\) is a sample from the distribution of \(X_l\), and, for any fixed \(n,\) the sample \(X_{l}^{(1,n)},\ldots,X_{l}^{(N_d,n)}\) is drawn from the conditional distribution of \(X_{l}\) given \(X_{l-1}=X_{l-1}^{(n)}\).
As an estimate for $V_0$ the random variable
$V_{N,N_d}$ is biased high (see~\eqref{eq:27082017a1} below).
The next theorem presents a bound for its
mean squared error (MSE).

\begin{thm}\label{berm:prop_vnm}
We have for the estimator $V_{N,N_d}$ 
\begin{equation}\label{eq:27082017a1}
\EE V_{N,N_d}\ge V_0,
\end{equation}
i.e.\ it is an upper bound for $V_0$
and hence for $V^*_0$. Moreover, it holds 
\begin{equation}\label{berm:eq:bound_vnm}
\begin{split}
&\EE\left[(V_{N,N_d}-V_0)^2\right]\\
&\le\frac4{N_d}\left(1+\frac1N\right)\sum_{l=1}^J \EE\left[\Var\left[v_{l}(X_{l}) | X_{l-1}\right]\right]
+\frac4N\sum_{l=1}^J \EE\left[\Var\left[v^*_l(X_l)-v_l(X_l) | X_{l-1}\right]\right]\\
&\le\frac4{N_d}\left(1+\frac1N\right)\sum_{l=1}^J \EE\left[\Var\left[v_{l}(X_{l}) | X_{l-1}\right]\right]
+\frac4N\sum_{l=1}^J
\EE\left[(v^*_l(X_l)-v_l(X_l))^2\right].
\end{split}
\end{equation}
\end{thm}

Statement~\eqref{eq:27082017a1} is known
(see Remark~3.2 in \cite{BSD})
and presented here only to make the exposition self-contained.
On the other hand, bound~\eqref{berm:eq:bound_vnm}
for the MSE of $V_{N,N_d}$ is new.
Several related quantities were extensively studied in
Chen and Glasserman \cite{chen2007additive}
and Belomestny et al \cite{BSD},
but neither of these papers contains such a bound.

Inequality~\eqref{berm:eq:bound_vnm}
is surprising because none of $V_{N,N_d}$
or $V_0$ involves the real price
$(v^*_l(X_l))$.
Conceptually, the real price comes into play
as it is inherent in the optimal stopping problem
for $(g_j(X_j))$
(notice that $(g_j(X_j))$ constitutes our problem data).
On the technical side, the real price appears in the proof
due to the almost sure property~\eqref{berm:eq:as_doob}.
More precisely, the random variable
$\max_{0\le j\le J}(g_j(X_j)-Y^*_j)$,
being degenerate, can be introduced
into~\eqref{eq:17012018a1}
without changing the variance term
in~\eqref{eq:17012018a1}.

\begin{remark}\label{rem:27082017a1}
(i) Bound~\eqref{berm:eq:bound_vnm} is very informative, as it not only gives an error estimate for \(V_{N,N_d},\) but also shows ways to improve the quality of $V_{N,N_d}$. While the second term in the r.h.s.\ of~\eqref{berm:eq:bound_vnm} can be reduced by making the  bound \(v_{l}\) closer to \(v^*_{l},\) the first one can be made smaller by reducing the magnitude of the conditional variances \(\Var\left[v_{l}(X_{l})|X_{l-1}\right].\)

(ii) In addition to the composite
bound~\eqref{berm:eq:bound_vnm} for the MSE
of $V_{N,N_d}$ it is instructive to see
what in this bound accounts for the squared bias
and what for the variance.
We will see in the proof of Theorem~\ref{berm:prop_vnm}
that
\begin{equation}\label{eq:29082017b1}
\left(\EE V_{N,N_d}-V_0\right)^2\le\frac{4}{N_d}\sum_{l=1}^J
\EE\left[\Var\left[v_l(X_l) | X_{l-1}\right]\right]
\end{equation}
and
\begin{equation}\label{eq:29082017b2}
\Var\left[V_{N,N_d}\right]
\le\frac1N\frac4{N_d}\sum_{l=1}^J \EE\left[\Var\left[v_{l}(X_{l}) | X_{l-1}\right]\right]
+\frac4N\sum_{l=1}^J \EE\left[\Var\left[v^*_l(X_l)-v_l(X_l) | X_{l-1}\right]\right].
\end{equation}
Roughly speaking, this means that $N$ (resp.\ $N_d$)
accounts for the variance (resp.\ the bias)
of the estimator $V_{N,N_d}$.
That is, we need to increase $N$ (resp.\ $N_d$)
in order to reduce the variance (resp.\ the bias).
On top of that we can observe a more delicate effect
that increasing $N_d$ alone (i.e.\ with $N$ being fixed)
also reduces a (small) part of the variance of $V_{N,N_d}$.
\end{remark}

Bound~\eqref{berm:eq:bound_vnm} for the MSE of $V_{N,N_d}$
also enables us to analyse the complexity
of the dual nested simulations approach.
Since the cost of computing \(V_{N,N_d}\) is of order \(N N_d\)
(recall that $J$ is fixed for now),
the overall complexity of the estimate \(V_{N,N_d}\),
i.e.\ the minimal cost needed to achieve
$\EE\left[(V_{N,N_d}-V_0)^2\right]\le\varepsilon^2$,
is of order $\varepsilon^{-4}$.
In the next two sections we will develop
a regression-based approach,
which will result in a significant reduction of the complexity
(see Remark~\ref{berm:rem}).

Let us mention  two relevant modifications of the nested dual algorithm  proposed in the literature. Firstly, in Belomestny et al \cite{belomestny2009true} an algorithm not involving sub-simulation was suggested, where an approximation for the Doob martingale  was constructed using the martingale representation theorem and some approximation of the true price process. 
However, that method requires an additional discretization of stochastic integrals and suffers from some instability for small discretization steps.
Secondly, a multilevel-type algorithm was developed
in~Belomestny et al \cite{BSD},
which has a similar performance, in terms of complexity,
as the algorithm presented in the next sections,
but works under very different conditions
(e.g.\ the algorithm in \cite{BSD}
does not take advantage of the smoothness properties
of the involved conditional expectations).
\par
In a similar way, a nested simulations estimator $U_{N,N_d}$
for $U_0$ of~\eqref{eq:upper_eep} can be  constructed as follows
$$
U_{N,N_d}=\frac{1}{N} \sum_{n=1}^N
\left[
g_J(X^{(n)}_{J})+\sum_{j=1}^{J}
(g_{j-1}(X^{(n)}_{j-1})-Z_{j,n,N_d})^{+}
\right],
$$
where
$$
Z_{j,n,N_d}=\frac{1}{N_d}\sum_{n_d=1}^{N_d} v_{j}(X^{(n_d,n)}_{j}),
\quad j=1,\ldots,J,
$$
where the sample \(X_{j}^{(1,n)},\ldots,X_{j}^{(N_d,n)}\) is drawn from the conditional distribution of \(X_{j}\) given \(X_{j-1}=X_{j-1}^{(n)}\).
In the next proposition, we present
a new bound for the MSE of \(U_{N,N_d}\).

\begin{prop}\label{prop:04032018a1}
We have for the estimator $U_{N,N_d}$
\begin{equation}\label{eq:04032018a1}
\EE U_{N,N_d}\ge U_0,
\end{equation}
i.e.\ it is an upper bound for $U_0$
and hence for $V^*_0$. Moreover, it holds 
\begin{equation}\label{eq:04032018a2}
\begin{split}
&\EE\left[(U_{N,N_d}-U_0)^2\right]\\
&\le\frac J{N_d}
\sum_{j=1}^J \EE\left[\Var\left[v_{j}(X_{j}) | X_{j-1}\right]\right]
+\frac1N\Var\left[
g_J(X^{(1)}_{J})+\sum_{j=1}^{J}
(g_{j-1}(X^{(1)}_{j-1})-Z_{j,1,N_d})^{+}
\right].
\end{split}
\end{equation}
\end{prop}

Again, bound~\eqref{eq:04032018a2}
shows a way of improving the quality
of the estimator $U_{N,N_d}$
by variance reduction technique:
the first term on the right-hand side can be made smaller
by reducing the magnitude of the conditional variances
\(\Var\left[v_{l}(X_{l})|X_{l-1}\right]\).
Recall that this also improves the quality of the dual
nested estimator $V_{N,N_d}$.
The second summand
on the right-hand side of~\eqref{eq:04032018a2}
is of order $J^2/N$
whenever all functions $g_j$,
$j=0,\ldots,J$, are uniformly bounded,
but this estimate $J^2/N$ is usually somewhat rough.
In specific situations the generic
bound~\eqref{eq:04032018a2}
should be complemented with
specific bounds for the second term
on the right-hand side of~\eqref{eq:04032018a2}.

\section{Variance reduction via regression}
\label{berm:sec:rbvr}
Usually the process $(X_t)_{t\in[0,T]}$
cannot be simulated exactly, and one has to use
some approximation of it.
Suppose that, for some \(\Delta>0,\) the time approximations $X_{\Delta,l\Delta},$ $l=0,\ldots,L,$ with \(L=\lfloor T/\Delta\rfloor\geq J\)
satisfy the following recurrence relations
\begin{equation}\label{berm:eq:01092016a1}
X_{\Delta,l\Delta}=\Phi_{l}(X_{\Delta,(l-1)\Delta},\xi_{l}),\quad l=1,\ldots,L,\quad X_{\Delta,0}=x_0,
\end{equation}
for some i.i.d.\ random vectors \(\xi_l\in \mathbb{R}^m\) with distribution \(\mu\)
and some Borel-measurable
functions $\Phi_{l}\colon\mathbb{R}^{d+m}\to\mathbb{R}^d.$
By $(\cG_l)_{l\in\{0,\ldots,L\}}$ we denote the filtration
with $\cG_0=\mathrm{triv}$
generated by $(\xi_l)_{l=1,\ldots,L}$.
It follows from~\eqref{berm:eq:01092016a1}
that $(X_{\Delta,l\Delta})_{l\in\{0,\ldots,L\}}$
is a $(\cG_l)$-Markov process.
Let \((\phi_k)_{k\in \mathbb{Z}_+}\) be a complete orthonormal system in \(L^2(\mathbb{R}^m,\mu)\) with \(\phi_0\equiv 1\). In particular,
\begin{eqnarray*}
\EE[\phi_i(\xi)\phi_j(\xi)]=\delta_{ij},\quad i,j\in  \mathbb{Z}_{+}.
\end{eqnarray*}
Notice that this implies that the random variables
$\phi_k(\xi)$, $k\ge1$, are centered.

The following result can be viewed as a
discrete-time analogue of the Clark-Ocone formula
or as a conditioned version of Theorem~2.1 in \cite{BHNU}.

\begin{thm}\label{berm:cor: expansion_main}
Consider some $j<p$ in $\{0,1,\dots,L\}$.
It holds for any Borel-measurable function \(f\) with \(\EE[\left|f(X_{\Delta,p\Delta})\right|^2]<\infty\)
\begin{equation}
f(X_{\Delta,p\Delta})=\EE\left[\left.f(X_{\Delta,p\Delta})\right|X_{\Delta,j\Delta}\right]+\sum_{k\geq1}\sum_{l=j+1}^{p }a_{p,l,k}(X_{\Delta,(l-1)\Delta})\phi_{k}(\xi_l),\label{berm:eq:expansioN_rond}
\end{equation}
where the series in the r.h.s.\ converges in $L^{2}$ sense. The coefficients
in~(\ref{berm:eq:expansioN_rond}) can be computed via 
\[
a_{p,l,k}(x)=\EE\left[\left.f(X_{\Delta,p\Delta})\phi_{k}\left(\xi_l\right)\right|X_{\Delta,(l-1)\Delta}=x\right]
\]
for $l\in\{j+1,\ldots,p\}$ and $k\in\mathbb N$.
\end{thm}

For fixed $j<p$ in $\{0,1,\dots,L\}$, define
\begin{eqnarray}
\label{berm:eq:hjp}
M_{j,p}=\sum_{k\geq1}\sum_{l=j+1}^{p }a_{p,l,k}(X_{\Delta,(l-1)\Delta})\phi_{k}(\xi_l)
\end{eqnarray}
and notice that
$\EE[M_{j,p}|\cG_j]=0$ a.s.\ and, in particular,
$\EE[M_{j,p}|X_{\Delta,j\Delta}]=0$ a.s.
Theorem~\ref{berm:cor: expansion_main} implies that  
\begin{eqnarray*}
\Var\left[f(X_{\Delta,p\Delta})-M_{j,p}|X_{\Delta,j\Delta}\right]=0
\quad a.s.,
\end{eqnarray*}
hence \(M_{j,p}\) is a perfect control variate for estimating
$\EE[f(X_{\Delta,p\Delta})|X_{\Delta,j\Delta}]$.
In order to use the control variate \(M_{j,p},\) we need to compute the coefficients \(a_{p,l,k}.\) This can be done by using regression in the following way: first we generate \(N_r\) discretized paths \(X^{(n)}_{\Delta,1\Delta},\ldots,X^{(n)}_{\Delta,L\Delta},\) \(n=N+1,\ldots,N+N_r,\) of the process \(X\)
(so-called ``training paths'')
and then solve the least squares optimization problems
\begin{eqnarray*}
\hat a_{p,l,k}=\arginf_{\psi\in \mathrm{span}(\psi_1,\ldots,\psi_Q)}\sum_{n=N+1}^{N+N_r} \left |f(X^{(n)}_{\Delta,p\Delta})\phi_{k}(\xi^{(n)}_l)-\psi(X^{(n)}_{\Delta,(l-1)\Delta})\right|^2,
\end{eqnarray*}
for \(l=j+1,\ldots,p,\) where \(\psi_1,\ldots,\psi_Q\) is a set of basis functions on \(\mathbb{R}^d.\) Furthermore, we truncate the summation in~\eqref{berm:eq:hjp} to get an implementable version of the control variate \(M_{j,p}\)
\begin{eqnarray}\label{berm:eq:01092016a2}
\hat M_{j,p,K}=\sum_{k=1}^K\sum_{l=j+1}^{p }\hat a_{p,l,k}(X_{\Delta,(l-1)\Delta})\phi_{k}(\xi_l).
\end{eqnarray}
To make clear how to understand~\eqref{berm:eq:01092016a2},
we remark that the random vectors $\xi_l$, $l=1,\ldots,L$,
in~\eqref{berm:eq:01092016a2}
are independent of the $N_r$ training paths $(X^{(n)}_{\Delta,l\Delta})$
used to obtain the regression-based estimates
$\hat a_{p,l,k}$,
while the (``testing'') path
$(X_{\Delta,l\Delta})$ in the argument of
$\hat a_{p,l,k}$ in~\eqref{berm:eq:01092016a2}
is constructed via those random vectors $\xi_l$
according to~\eqref{berm:eq:01092016a1}
(and hence is independent of the training paths).

Let us note that
$\EE\left[\hat M_{j,p,K}|X_{\Delta,j\Delta}\right]=0$
due to the martingale transform structure
in~\eqref{berm:eq:01092016a2}
(recall that $\EE\phi_k(\xi_l)=0$ for $k\ge1$),
i.e.\ $\hat M_{j,p,K}$ is indeed a valid control variate
in that it does not introduce any bias.
The properties
of such a control variate
are summarised  in the following theorem.

\begin{thm}\label{berm:theorem:regression_cv}
Consider some $j<p$ in $\{0,1,\dots,L\}$.
Suppose that the function \(f\) is uniformly bounded
by a constant~\(F\).
By $\tilde a_{p,l,k}
$ we denote the truncated
at the level $F$ estimate\footnote{To explain this truncation we notice that $|a_{p,l,k}(x)|\le F$ for all~$x$ by the assumption.}
\begin{equation}\label{berm:eq:08092016a1}
\tilde a_{p,l,k}(x)=T_F \hat a_{p,l,k}(x)
=\begin{cases}
\hat a_{p,l,k}(x)&\text{if }|\hat a_{p,l,k}(x)|\le F,\\
F\sgn\hat a_{p,l,k}(x)&\text{otherwise},
\end{cases}
\end{equation}
and by $\tilde M_{j,p,K}$ the control variate
defined like in~\eqref{berm:eq:01092016a2} but with
$\hat a_{p,l,k}$ replaced by $\tilde a_{p,l,k}$.
Furthermore, assume that, for some \(\beta\geq 0\) and \(B_\beta>0\),
\begin{eqnarray}\label{berm:ass}
\sum_{k=1}^\infty k^\beta\,
\sum_{l=j+1}^p
\EE[a^2_{p,l,k}(X_{\Delta,(l-1)\Delta})]\leq B_\beta
\end{eqnarray}
and the set of basis functions \(\psi_1,\ldots,\psi_Q\) is chosen in such a way that, for all $k\in\NN$,
\begin{eqnarray}\label{eq:29082017a4}
\sum_{l=j+1}^{p}\,
\inf_{\psi\in\mathrm{span}(\psi_{1},\ldots,\psi_{Q})}
\EE\left[\left|a_{p,l,k}(X_{\Delta,(l-1)\Delta})-\psi(X_{\Delta,(l-1)\Delta})\right|^{2}\right]\leq D_\kappa Q^{-\kappa},
\end{eqnarray}
for some constants \(\kappa\geq 0\) and \(D_\kappa>0\).
Then
\begin{align}
\EE\left[\Var\left[\left.
f(X_{\Delta,p\Delta})-\tilde M_{j,p,K}
\right|X_{\Delta,j\Delta}\right]\right]
&\leq
\tilde c F^2 (p-j) K \frac{Q(\log(N_r)+1)}{N_r}
\label{eq:29082017a1}\\
&\phantom{\leq}+8 D_\kappa K Q^{-\kappa}+B_\beta K^{-\beta}
\notag
\end{align}
with some universal constant~$\tilde c$.
\end{thm}

\begin{remark}\label{rem:29082017a1}
Theorem~\ref{berm:theorem:regression_cv}
simplifies in the case when the random vectors $\xi_l$
are discrete with finite number of atoms
(e.g.\ think about a weak approximation of an SDE
via scheme~\eqref{berm:eq:01092016a1} with discrete
random vectors $\xi_l$;
see Part~VI in \cite{KP}
or Chapter~2 in \cite{MilsteinTretyakov:2004}).
In this case, the space $L^2(\mathbb R^m,\mu)$
is finite-dimensional, and hence the basis
$(\phi_k)_{k\in\{0,1,\ldots,K_{max}\}}$
is finite consisting of say $K_{max}+1$ elements
(recall that $\phi_0\equiv1$).
Then we can drop assumption~\eqref{berm:ass},
while the conclusion~\eqref{eq:29082017a1}
can be replaced with
$$
\EE\left[\Var\left[\left.
f(X_{\Delta,p\Delta})-\tilde M_{j,p,K_{max}}
\right|X_{\Delta,j\Delta}\right]\right]
\leq
\tilde c F^2 (p-j) K_{max} \frac{Q(\log(N_r)+1)}{N_r}
+8 K_{max} D_\kappa Q^{-\kappa}.
$$
\end{remark}

\section{Dual upper bounds with reduced complexity}
\label{berm:sec:rbdub}
Next we apply the results of the previous section to the nested simulations of dual upper bounds.
For the sake of clarity assume that the exercise times coincide with the discretization time grid for some \(\Delta>0\), i.e.\ \(L=J\).
Instead of $V_0$,
which is constructed in~\eqref{berm:eq:repr_upper}
via the exact process,
we are now going to estimate
its analogue $V_{\Delta,0}$
constructed via the discretized process
\begin{equation}\label{berm:eq:30092016a1}
V_{\Delta,0}=\EE\left[ \max_{0\leq j\leq J}\left(
g_{j}(X_{\Delta,j\Delta})-Y_{\Delta,j\Delta} \right) \right]
\end{equation}
with \(Y_{\Delta,j\Delta}=\sum_{l=1}^{j}\left( v_{l}(X_{\Delta,l\Delta})-\EE\left[ v_{l}(X_{\Delta,l\Delta})|
X_{\Delta,(l-1)\Delta}\right] \right)\).
For any  \(j=1,\ldots,J,\) we need to compute the conditional expectations \(\EE\left[ v_{j}(X_{\Delta,j\Delta})|X_{\Delta,(j-1)\Delta}\right].\)
By Theorem~\ref{berm:cor: expansion_main},
we have the following representation
\begin{eqnarray}
\label{berm:eq:repr_vj}
v_j(X_{\Delta,j\Delta})=\EE\left[\left.v_j(X_{\Delta,j\Delta})\right|X_{\Delta,(j-1)\Delta}\right]+\sum_{k\geq1} a_{j,k}(X_{\Delta,(j-1)\Delta})\phi_{k}(\xi_j),
\end{eqnarray}
where 
\begin{eqnarray}
\label{berm:eq:coeff_dual}
a_{j,k}(x)=\EE\left[\left.v_j(X_{\Delta,j\Delta})\phi_{k}\left(\xi_j\right)\right|X_{\Delta,(j-1)\Delta}=x\right],
\end{eqnarray}
provided \(\EE\left[v^2_j(X_{\Delta,j\Delta})\right]<\infty.\) Representation~\eqref{berm:eq:repr_vj} implies that 
\begin{align}
\label{berm:var_v}
\Var[v_{j}(X_{\Delta,j\Delta})-M_j|X_{\Delta,(j-1)\Delta}]=0\quad a.s.
\end{align}
for 
\begin{align}
\label{berm:cv_M}
M_j=\sum_{k\geq1} a_{j,k}(X_{\Delta,(j-1)\Delta})\phi_{k}(\xi_j).
\end{align}
The control variates \(M_1,\ldots,M_J\) cannot be used directly, since the coefficients \(a_{j,k}\) are unknown
and we need to truncate the summation in~\eqref{berm:cv_M}
to get an implementable quantity.
Given that we can find implementable approximations
for $M_j$, say $\hat M_j$,
satisfying $\EE\left[\hat M_j|X_{\Delta,(j-1)\Delta}\right]=0$,
the idea is now to use the random variables
$v_j(X_{\Delta,j\Delta})-\hat M_j$
in the nested simulations step to approximate
$\EE\left[v_j(X_{\Delta,j\Delta})|X_{\Delta,(j-1)\Delta}\right]$.
Indeed, $\Var[v_{j}(X_{\Delta,j\Delta})-\hat M_j|X_{\Delta,(j-1)\Delta}]$
will be close to zero for good approximations $\hat M_j$
(cf.~\eqref{berm:var_v}).

So first we estimate the coefficients \(a_{l,k}\) by a preliminary regression using \(N_r\) discretized paths of the process \(X\) and \(Q\) basis functions (see Section~\ref{berm:sec:rbvr}).
In this way we construct the approximation of the control variate \(M_l\) given by 
\begin{eqnarray}\label{berm:eq:08092016a2}
\hat M_{l,K}=\sum_{k=1}^K \hat a_{l,k}(X_{\Delta,(l-1)\Delta})\phi_{k}(\xi_l).
\end{eqnarray}
Now fix some natural numbers \(N_d,\) \(N\) and consider the  dual estimate 
\begin{align}\label{berm:eq:08092016a5}
\hat V_{N,N_d,K}=\frac{1}{N}\sum_{n=1}^N\left[ \max_{0\leq j\leq J}\left(
g_{j}(X^{(n)}_{\Delta,j\Delta})-\hat Y_{j,n,N_d,K} \right) \right],
\end{align}
where 
\begin{eqnarray}\label{berm:eq:08092016a4}
\hat Y_{j,n,N_d,K}=\sum_{l=1}^{j}\left(v_{l}(X^{(n)}_{\Delta,l\Delta})-\frac{1}{N_d}\sum_{n_d=1}^{N_d} \left(v_{l}(X^{(n_d,n)}_{\Delta,l\Delta})-\hat M_{l,K}^{(n_d,n)}\right) \right)
\end{eqnarray}
with 
\begin{eqnarray}\label{berm:eq:08092016a3}
\hat M^{(n_d,n)}_{l,K}=\sum_{k=1}^K \hat a_{l,k}(X^{(n)}_{\Delta,(l-1)\Delta})\phi_{k}(\xi_l^{(n_d,n)}).
\end{eqnarray}
We now can prove the following result.

\begin{thm}\label{berm:prop:29092016a1}
Assume that all functions \(v_j,\) \(j=1,\ldots,J,\) are uniformly bounded by a constant~\(F\).
By $\tilde a_{j,k}$ we denote the truncated
at the level $F$ estimate
defined as in~\eqref{berm:eq:08092016a1},
and by $\tilde M_{l,K}$
(resp.\ $\tilde M^{(n_d,n)}_{l,K}$,
$\tilde Y_{j,n,N_d,K}$,~$\tilde V_{N,N_d,K}$)
the quantities defined like
in~\eqref{berm:eq:08092016a2}
(resp.\ \eqref{berm:eq:08092016a3},
\eqref{berm:eq:08092016a4},~\eqref{berm:eq:08092016a5})
but with ``hats'' replaced by ``tildes''.
Suppose that the coefficients \((a_{j,k})\) defined
in~\eqref{berm:eq:coeff_dual} satisfy, for all $j=1,\ldots,J$,
\begin{eqnarray}\label{berm:eq:29092016a1}
\sum_{k=1}^\infty k^\beta\,\EE[a^2_{j,k}(X_{\Delta,(j-1)\Delta})]\leq B_\beta
\end{eqnarray}
with some \(\beta\geq 0\) and \(B_\beta>0\)
and that the basis functions
\(\psi_1,\ldots,\psi_Q\) are chosen in such a way that,
for all $j=1,\ldots,J$ and $k\in\mathbb N$,
\begin{eqnarray}
\label{berm:eq:rho}
\inf_{\psi\in\mathrm{span}(\psi_{1},\ldots,\psi_{Q})}\EE\left[\left|a_{j,k}(X_{\Delta,(j-1)\Delta})-\psi(X_{\Delta,(j-1)\Delta})\right|^{2}\right]\leq D_\kappa Q^{-\kappa}
\end{eqnarray}
with some \(\kappa\geq 0\) and \(D_\kappa>0.\) Then it holds 
\begin{align}
\EE\left[(\tilde V_{N,N_d,K}-V_{\Delta,0})^2\right]
&\leq\frac{4J}{N_d}\left(1+\frac{1}{N}\right)
\left[\tilde c F^2 K \frac{Q(\log(N_r)+1)}{N_r}
+8 D_\kappa K Q^{-\kappa}+B_\beta K^{-\beta}\right]
\notag\\
&\phantom{=}+\frac4N
\sum_{l=1}^{J}\EE\left[\left(v_{l}^{*}(X_{\Delta,l\Delta})-v_{l}(X_{\Delta,l\Delta})\right)^{2}\right]
\label{berm:eq:bound_vnm_regr}
\end{align}
with some universal constant~$\tilde c$.
\end{thm}

Let us notice that the statement similar to that
in Remark~\ref{rem:29082017a1}
applies here as well.


\subsection{Complexity analysis for fixed~$J$}
Theorem~\ref{berm:prop:29092016a1} allows us
to carry out complexity analysis of our algorithm.
First note that the overall cost of computing the estimator $\tilde V_{N,N_d,K}$ is of order 
\begin{align}
\label{berm:cost_algorithm}
JK\max\left\{N_rQ^2,NQ,NN_d\right\},
\end{align}
where the first term in~\eqref{berm:cost_algorithm} comes from the computation of the regression coefficients, the second one from the computation of $\tilde a_{l,k}(X^{(n)}_{\Delta,(l-1)\Delta})$ and the last one from the computation of $\tilde M^{(n_d,n)}_{l,K}$
(other terms involved in the computation are dominated by one of these quantities).
Given $\beta>0$ and $\kappa>0$
as in Theorem~\ref{berm:prop:29092016a1},
we have the following constraints
\begin{align}
\label{berm:mse_constr}
\max\left\{\frac{JKQ\log(N_r)}{N_rN_d},\frac{JB_\beta}{K^\beta N_d},\frac{JD_\kappa K}{ Q^{\kappa}N_d},\frac{J}{N}\right\}\lesssim\varepsilon^2
\end{align} 
to ensure the condition $\EE\left[|\tilde V_{N,N_d,K}-V_{\Delta,0}|^2\right]\lesssim\varepsilon^2$.

Notice that we are interested in getting
the order of complexity in $\varepsilon$
as $\varepsilon\searrow0$.
To this end, we need to determine the parameters
$N$, $N_r$, $N_d$, $K$ and $Q$ via $\varepsilon$
in such a way that the order of complexity
of $\tilde V_{N,N_d,K}$
(given by~\eqref{berm:cost_algorithm})
is minimal under the constraint~\eqref{berm:mse_constr}.
Since $B_\beta$, $D_\kappa$ and $J$ are constants,
they can be dropped
from~\eqref{berm:cost_algorithm} and~\eqref{berm:mse_constr}.
Straightforward but lengthy calculations\footnote{For more detail, see Section~\ref{sec:append_compl}.} now show that the overall complexity  of  \(\tilde V_{N,N_d,K}\) is bounded from above by
\begin{equation}
\label{berm:compl_sol}
C_{J,\beta,\kappa}\,\varepsilon^{-\frac{4(\beta+1)(\kappa+3)+4\kappa}{(\beta+1)(\kappa+3)+\beta\kappa}}\sqrt{|\log\varepsilon|},
\end{equation}
where the constant $C_{J,\beta,\kappa}$
does not depend on~$\varepsilon$.
Moreover, the dependence structure in $C_{J,\beta,\kappa}$
on the parameters $\beta$, $\kappa$ and $J$
is given by the formula
$C_{J,\beta,\kappa}=c J^2 B_\beta^{3/(1+\beta)} D_\kappa^{3/(3+\kappa)}$
with some universal constant~$c$.
We, finally, discuss the complexity estimate~\eqref{berm:compl_sol}.

\begin{remark}\label{berm:rem}
(i) We require to choose $\beta>1$ in order
to be better than the standard nested simulations approach
discussed in Section~\ref{berm:sec:ns}
because $\frac{4(\beta+1)(\kappa+3)+4\kappa}{(\beta+1)(\kappa+3)+\beta\kappa}<4$ whenever $\beta>1$.

(ii) We can achieve the complexity order
$\varepsilon^{-2-\delta}$, for arbitrarily small $\delta>0$,
whenever the parameters $\beta$ and $\kappa$
are sufficiently large.

(iii) In the limiting case \(\kappa=0,\) i.e., if the approximation error in \eqref{berm:eq:rho} does not converge to \(0\) (e.g.\ due to an inappropriate choice of basis functions), we end up with the complexity  of the standard nested approach of order $\varepsilon^{-4}$.
\end{remark}

In the next subsection
we present the complexity analysis
for the case of an increasing number of exercise dates $J\to\infty$.
We  also  take the discretization error into account,
which is the order (in~$J$, $J\to\infty$)
of the difference between the upper bound $V_0$
for the (continuous time) American option price
and the upper bounds $V_{\Delta,0}$
for the Bermudan option prices with $\Delta=T/J$.

\subsection{Complexity analysis for $J\to\infty$}
\label{berm:sec:complan2}
To approximate an upper bound $V_0$
for a true American (rather than Bermudan) option,
we now let $J$ tend to infinity.
We shall compare the complexities of the
\emph{standard approach}
(the one of Section~\ref{berm:sec:ns} applied to the discretized process)
and of the
\emph{regression-based approach}
(the one described in the beginning of Section~\ref{berm:sec:rbdub}).

\medskip
\emph{Standard approach:}
Set $\Delta=T/J,$ then the estimate for $V_{\Delta,0}$ of~\eqref{berm:eq:30092016a1} is
$$
V_{\Delta,N,N_d}=\frac{1}{N}\sum_{n=1}^N\left[ \max_{0\leq j\leq J}\left(
g_{j}(X^{(n)}_{\Delta,j\Delta})-Y_{\Delta,j\Delta,n,N_d} \right) \right],
$$
where 
\begin{eqnarray*}
Y_{\Delta,j\Delta,n,N_d}=\sum_{l=1}^{j}\left(v_{l}(X^{(n)}_{\Delta,l\Delta})-\frac{1}{N_d}\sum_{n_d=1}^{N_d} v_{l}(X^{(n_d,n)}_{\Delta,l\Delta}) \right),
\quad j=0,\ldots,J.
\end{eqnarray*}
The analogue of~\eqref{berm:eq:bound_vnm} takes the form
\begin{align}
\notag
\EE\left[|V_{\Delta,N,N_d}-V_{\Delta,0}|^2\right]\leq&
\frac{4\sum_{l=1}^{J}\EE\left[\Var\left[v_{l}(X_{\Delta,l\Delta})|X_{\Delta,(l-1)\Delta}\right]\right]}{N_d}\left(1+\frac{1}{N}\right)\\
\label{berm:eq:30092016a2}
&+\frac{4\sum_{l=1}^{J}\EE\left[\left|v_{l}^{*}(X_{\Delta,l\Delta})-v_{l}(X_{\Delta,l\Delta})\right|^{2}\right]}{N}.
\end{align}
Since we are considering American
options in this section, the estimate $V_{\Delta,N,N_d}$
can be viewed as an estimate for $V_0$ rather than
for $V_{\Delta,0}$, i.e.\ this is
$\EE\left[|V_{\Delta,N,N_d}-V_0|^2\right]$
that should be of order $\varepsilon^2$
in the complexity analysis.
Therefore, we need an assumption about the order
of the discretization error $V_{\Delta,0}-V_0$.
It seems reasonably general to assume that
it is of order $\frac1{\sqrt J}$.
However, the discretization error might
be of a different order in specific situations (see \cite{dupuis2005convergence}).
That is why we impose a more general assumption:

\begin{enumerate}
\item[(A1) ] $\,V_{\Delta,0}-V_0$ is of order
$J^{-\alpha}$ as $J\to\infty$
with some $\alpha>0$.
\end{enumerate}

\noindent
We also need an assumption on the order
of the second term in the right-hand side
of~\eqref{berm:eq:30092016a2}
(which is also present in~\eqref{berm:eq:bound_vnm_regr}):

\begin{enumerate}
\item[(A2) ] $\,\sum_{l=1}^{J}\EE\left[\left|v_{l}^{*}(X_{\Delta,l\Delta})-v_{l}(X_{\Delta,l\Delta})\right|^{2}\right]$ is of order
$J^q$ as $J\to\infty$
with some $q\in[0,1]$.
\end{enumerate}

\noindent
A typical-to-expect situation here is $q=1$.
Another interesting variant is $q=0$:
here the strategy 
is to use better and better approximations $v_l$
for $v^*_l$ at each time point $l=1,\ldots,J,$
as $J$ grows (see, e.g., Zanger \cite{zanger2013quantitative} for bounds on  
\(\EE\left[\|v_{l}^{*}-v_{l}\|^2\right].\))
Finally, as for the first term on the right-hand side
of~\eqref{berm:eq:30092016a2}
it is reasonable to assume only that

\begin{enumerate}
\item[(A3) ] $\,\sum_{l=1}^{J}\EE\left[\Var\left[v_{l}(X_{\Delta,l\Delta})|X_{\Delta,(l-1)\Delta}\right]\right]$ is of order
$J$ as $J\to\infty.$
\end{enumerate}

\noindent
The overall cost of computing the estimate
$V_{\Delta,N,N_d}$ is of order $JN_dN$.
Thus, we need to minimize this cost order
under the constraint
$$
\max\left\{\frac1{J^{2\alpha}},
\frac J{N_d},\frac{J^q}N\right\}\lesssim\varepsilon^2,
$$
which ensures that
$\EE\left[|V_{\Delta,N,N_d}-V_0|^2\right]\lesssim\varepsilon^2$
(see~\eqref{berm:eq:30092016a2} and (A1)--(A3)).
This leads to the complexity of $V_{\Delta,N,N_d}$
of order $\varepsilon^{-4-\frac{2+q}\alpha}$.
For instance, in the case $\alpha=1/2$, $q=1$
(resp.\ $\alpha=1/2$, $q=0$)
we get the complexity
$O(\varepsilon^{-10})$
(resp.~$O(\varepsilon^{-8})$).

\medskip
\emph{Regression-based approach:}
We suppose that the assumptions
of Theorem~\ref{berm:prop:29092016a1}
are satisfied uniformly in $J\in\mathbb N$
and again assume (A1) and~(A2)
(as for~(A3), we do not need it here).
The cost of computing $\tilde V_{N,N_d,K}$ is of order
$$
JK\max\left\{N_rQ^2,NQ,NN_d\right\}.
$$
We need to minimize this under the constraints
$$
\max\left\{\frac1{J^{2\alpha}},
\frac{JKQ\log(N_r)}{N_rN_d},\frac{JB_\beta}{K^\beta N_d},\frac{JD_\kappa K}{ Q^{\kappa}N_d},\frac{J^q}{N}\right\}\lesssim\varepsilon^2,
$$
which ensures that
$\EE\left[|\tilde V_{N,N_d,K}-V_0|^2\right]\lesssim\varepsilon^2$
(see~\eqref{berm:eq:bound_vnm_regr} and (A1)--(A2)).
Straightforward but lengthy calculations\footnote{The calculations are similar to the derivation of~\eqref{berm:compl_sol}, which is discussed in Section~\ref{sec:append_compl}.} show that
the overall complexity of $\tilde V_{N,N_d,K}$ is bounded from above by
\begin{equation}
\label{berm:compl_sol_J}
C_{\beta,\kappa}\,\varepsilon^{-\frac{(4\alpha+2+q)(\beta+1)(\kappa+3)+(\beta+4\alpha+1+q)\kappa}{\alpha(\beta+1)(\kappa+3)+\alpha\beta\kappa}}\sqrt{|\log\varepsilon|},
\end{equation}
where the constant $C_{\beta,\kappa}$
does not depend on~$\varepsilon$.
Moreover, the dependence on $\beta$ and $\kappa$
is described by the formula
$C_{\beta,\kappa}=cB_\beta^{3/(1+\beta)} D_\kappa^{3/(3+\kappa)}$
with some universal constant~$c$.
We, finally, discuss the complexity estimate~\eqref{berm:compl_sol_J}.

\begin{remark}
(i) We again require to choose $\beta>1$
in order to be better than the standard approach
discussed above, because,
as a straightforward calculation shows,
$$\frac{(4\alpha+2+q)(\beta+1)(\kappa+3)+(\beta+4\alpha+1+q)\kappa}{\alpha(\beta+1)(\kappa+3)+\alpha\beta\kappa}<4+\frac{2+q}\alpha$$
whenever $\beta>1$.

(ii) We can achieve the complexity order
$\varepsilon^{-2-\frac{3+q}{2\alpha}-\delta}$,
for arbitrarily small $\delta>0$,
whenever the parameters $\beta$ and $\kappa$
are sufficiently large.
In particular, this gives us
$O(\varepsilon^{-6-\delta})$
(resp.~$O(\varepsilon^{-5-\delta})$)
when $\alpha=1/2$, $q=1$
(resp.\ $\alpha=1/2$, $q=0$),
which is to be compared with
$O(\varepsilon^{-10})$
(resp.~$O(\varepsilon^{-8})$)
in the case of the standard approach.
\end{remark}

\section{Examples and discussion of conditions}
\label{berm:sec:examples}
Suppose that the process $(X_{t})_{t\in[0,T]}$ solves the SDE
\[
dX_{t}=\mu(X_{t})\,dt+\sigma(X_{t})\,dW_{t},\quad t\in[0,T],
\]
where $\mu$ and $\sigma$ are globally
Lipschitz functions $\mathbb R\to\mathbb R$.
Consider the Euler discretization scheme,
which is of the form 
\[
X_{\Delta,j\Delta}=X_{\Delta,(j-1)\Delta}+\mu(X_{\Delta,(j-1)\Delta})\,\Delta+\sigma(X_{\Delta,(j-1)\Delta})\,\xi_{j}\,\sqrt{\Delta},\quad j=1,\ldots,J,
\]
where $\xi_{1},\ldots,\xi_{J}$ are independent $N(0,1)$
random variables. In this case, we have
\[
\Phi_{\Delta}(x,y)=x+\mu(x)\,\Delta+\sigma(x)\,y\,\sqrt{\Delta},
\]
and the orthonormal system $(\phi_k)_{k\in\mathbb Z_+}$
in $L^2\left(\mathbb R,N(0,1)\right)$
can be chosen to be the system of normalised Hermite polynomials:
$$
\phi_k=\frac{H_k}{\sqrt{k!}},\quad
H_k(x)=(-1)^k e^{x^2/2} \frac{d^k}{dx^k} e^{-x^2/2}.
$$
Then the coefficients $a_{j,k}$ are given by formula
\begin{equation}\label{berm:eq:29092016b1}
a_{j,k}(x) =\frac{1}{\sqrt{k!}}\EE\left[v_j\left(x+\mu(x)\,\Delta+\sigma(x)\,\xi\,\sqrt{\Delta}\right)H_{k}(\xi)\right]
\end{equation}
with $\xi\sim N(0,1)$.
To get more insight into the behaviour of $a_{j,k}$ in~$k$,
we need to know the structure of the approximations~$v_j$.
While we did not assume anything on their structure until now,
in practice one often models $v_j$
as linear combinations of some basis functions,
e.g.\ polynomials
(for instance, in the Longstaff-Schwartz algorithm with a polynomial basis).
Let us now verify the assumption~\eqref{berm:eq:29092016a1}
in a couple of particular examples.

\begin{example}
Let
$$
v_j(y)=\sum_{i=0}^p \alpha_{j,i} y^i,\quad j=1,\ldots,J
$$
(think of polynomial basis functions).
Since, with $\xi\sim N(0,1)$,
$H_k(\xi)$ is orthogonal in $L^2$
to all polynomials in $\xi$
of degree less than~$k$,
it follows from~\eqref{berm:eq:29092016b1} that
$$
a_{j,k}\equiv0\quad\text{whenever }k\ge p+1.
$$
Then, for any $\beta>0$, there is an appropriate constant
$B_\beta>0$ such that, for all $j=1,\ldots,J$,
$$
\sum_{k=1}^\infty
k^\beta\,\EE[a_{j,k}^2(X_{\Delta,(j-1)\Delta})]
=\sum_{k=1}^p
k^\beta\,\EE[a_{j,k}^2(X_{\Delta,(j-1)\Delta})]
\le B_\beta.
$$
(Notice that, since the coefficients $\mu$ and $\sigma$
of the SDE are globally Lipschitz,
all polynomial moments of the Euler discretization
are finite, hence all
$\EE\left[a_{j,k}^2(X_{\Delta,(j-1)\Delta})\right]$
are finite.)
Thus, assumption~\eqref{berm:eq:29092016a1} is satisfied
and, moreover, we can take arbitrarily large $\beta>0$
(at a cost of possibly getting large~$B_\beta$).
\end{example}

\begin{example}
Let now
$$
v_j(y)=\sum_{l=-p}^p
\alpha_{j,l}\exp\{ihly\},
\quad j=1,\ldots,J,
$$
that is, at each time step $j=1,\ldots,J$
our approximations $v_j$ are trigonometric polynomials
with period $2\pi/h$, for some given $h>0$.
With $\xi\sim N(0,1)$ we have
\begin{align*}
a_{j,k}(x)
&=\frac{1}{\sqrt{k!}}
\EE\left[v_j\left(x+\mu(x)\Delta+\sigma(x)\sqrt{\Delta}\,\xi\right)H_{k}(\xi)\right]\\
&=\frac{1}{\sqrt{k!}}
\sum_{l=-p}^{p}
\alpha_{j,l}
\exp\left\{ihl(x+\mu(x)\Delta)\right\}
\EE\left[
\exp\left\{ihl\sigma(x)\sqrt{\Delta}\,\xi\right\}
H_k(\xi)\right].
\end{align*}
Using the definition of the Hermite polynomials
and integrating by parts $k$ times, we compute
$$
\EE\left[\exp\{ia\xi\}H_k(\xi)\right]
=(ia)^k \exp\left\{-\frac{a^2}2\right\}.
$$
Hence,
$$
|a_{j,k}(x)|
\le\frac{h^k\Delta^{k/2}}{\sqrt{k!}}
\sum_{l=-p}^{p}
|\alpha_{j,l}|
\,l^k\,
|\sigma(x)|^k
\exp\left\{-\frac{h^{2}l^{2}\sigma^2(x)\Delta}2\right\}.
$$
Assuming for simplicity that $\sigma$ is bounded,
we get
$$
|a_{j,k}(x)|\le\sqrt{\frac{K_j\,C^k}{k!}}
$$
with some positive constants $K_j$ and $C$.
Hence, for any $\beta>0$ and for all $j=1,\ldots,J$,
$$
\sum_{k=1}^\infty
k^\beta\,\EE[a_{j,k}^2(X_{\Delta,(j-1)\Delta})]
\le
\left[\max_{j=1,\ldots,J}K_j\right]
\sum_{k=1}^\infty
\frac{k^\beta\,C^k}{k!}
=: B_\beta<\infty.
$$
Thus, provided $\sigma$ is bounded,
for arbitrarily large $\beta>0$,
there exists an appropriate $B_\beta>0$ such that
assumption~\eqref{berm:eq:29092016a1} is satisfied.
\end{example}

\section{Numerical results}
\label{berm:sec:numerics}
As can be easily seen, the optimal solution for the parameter $N$ is of the same order (w.r.t.~$\varepsilon$)
both in the standard and in the regression-based approaches.
Therefore, let us ignore the error term
\begin{align}
\label{berm:error_N}
\frac4N
\sum_{l=1}^{J}\EE\left[\left|v_{l}^{*}(X_{\Delta,l\Delta})-v_{l}(X_{\Delta,l\Delta})\right|^{2}\right]
\end{align}
in~\eqref{berm:eq:bound_vnm_regr} and~\eqref{berm:eq:30092016a2}. Hence, we are interested in the remaining terms
\begin{equation}\label{berm:eq:02112016a1}
\EE\left[\Var\left[v_{l}(X_{\Delta,l\Delta})|X_{\Delta,(l-1)\Delta}\right]\right]
\end{equation}
and
\begin{equation}\label{berm:eq:02112016a2}
\EE\left[\Var\left[
v_l(X_{\Delta,l\Delta})-\tilde M_{l,K} | X_{\Delta,(l-1)\Delta}
\right]\right],
\end{equation}
for $l=1,\ldots,J$, respectively. In terms of the numerical implementation, we will choose $N$ large enough so that~\eqref{berm:error_N} does not really affect the overall error.
That is, we now consider $J$ and $N$ as fixed parameters.

\medskip
\emph{Standard approach with fixed $J$ and~$N$:}
We recall that the overall cost of computing the estimator
$V_{\Delta,N,N_d}$ is of order $JN_dN$.
Since we consider only the variance terms, we set $N_d\asymp\varepsilon^{-2}$ to ensure that (see~\eqref{berm:eq:30092016a2})
\begin{align}
\label{berm:condition_M}
\frac{4}{N_d}\left(1+\frac{1}{N}\right)\sum_{l=1}^{J}\EE[\Var\left[v_{l}(X_{\Delta,l\Delta})|X_{\Delta,(l-1)\Delta}\right]]\lesssim\varepsilon^2.
\end{align}
Thus, we have for the complexity
\begin{equation}\label{berm:eq:02112016a3}
\mathcal C_{standard}\asymp JN_dN\asymp\varepsilon^{-2}.
\end{equation}

\medskip
\emph{Regression-based approach with fixed $J$ and~$N$:}
The overall cost of computing the estimator $\tilde V_{N_d,N,K}$ is of order
\begin{align}
\label{berm:cost_algorithm_N}
JK\max\left\{N_rQ^2,NN_d\right\}.
\end{align}
Notice that, since $N$ is considered to be fixed,
the term $NQ$ (cf.~\eqref{berm:cost_algorithm})
is dominated by $N_rQ^2$.
We have the constraints
\begin{align}
\label{berm:mse_constr_N}
\max\left\{\frac{JKQ\log(N_r)}{N_rN_d},\frac{JB_\beta}{K^\beta N_d},\frac{JD_\kappa K}{ Q^{\kappa}N_d}\right\}\lesssim\varepsilon^2
\end{align} 
to ensure the condition
\begin{align}
\frac{4}{N_d}\left(1+\frac{1}{N}\right)\sum_{l=1}^J\EE\left[\Var\left[
v_l(X_{\Delta,l\Delta})-\tilde M_{l,K} | X_{\Delta,(l-1)\Delta}
\right]\right]\lesssim\varepsilon^2.
\end{align}
Then, the resulting complexity bound is given by
\begin{equation}\label{berm:eq:02112016a4}
\mathcal{C}_{regression}\lesssim C_{J,N,\beta,\kappa}\,\varepsilon^{-\frac{2(\beta+1)(\kappa+3)+2\kappa}{(\beta+1)(\kappa+3)+\beta\kappa}}\sqrt{|\log\varepsilon|},
\end{equation}
where $C_{J,N,\beta,\kappa}=cJ^{3/2}N^{1/2}B_\beta^{3/(1+\beta)} D_\kappa^{3/(3+\kappa)}$ with some universal constant~$c$.
Notice that the complexity in~\eqref{berm:eq:02112016a4}
is better than that in~\eqref{berm:eq:02112016a3}
whenever $\beta>1$.
Moreover, we can achieve the complexity order
$\varepsilon^{-1-\delta}$ in~\eqref{berm:eq:02112016a4},
for arbitrarily small $\delta>0$,
whenever the parameters $\beta$ and $\kappa$
are sufficiently large.


In constructing the numerical experiments below,
for the regression-based approach,
we need to choose several values of $\varepsilon$
and the values of $N_r$, $N_d$, $K$ and $Q$
for each value of~$\varepsilon$.
To this end, we use the ``limiting formulas''
as $\beta,\kappa\to\infty$.
Ignoring the remaining constants
as well as the log-term for $N_r$,
those ``limiting formulas'' give us
$N_r=O(\varepsilon^{-1})$,
$N_d=O(\varepsilon^{-1})$,
$K=O(1)$ and $Q=O(1)$.
In more detail,
we choose the parameters
for each $\varepsilon=2^{-i}$,
$i\in\left\{2,3,4,5,6\right\}$,
as follows:
$$
N=5\cdot 10^4,\quad K=1,\quad Q=d+2,\quad N_d=8\cdot\varepsilon^{-1},\quad N_r=256\cdot\varepsilon^{-1}.
$$
As for the basis functions, we use polynomials
of $d$ variables up to degree~1 as well as the function~$g_j$,
which will be independent of~$j$
(payoff of a Bermudan max-call option).
Altogether $Q=d+2$ basis functions.
Regarding the standard approach, we choose for each $\varepsilon=2^{-i}$, $i\in\left\{2,3,4,5\right\}$, the parameters via
$$
N=5\cdot 10^4,\quad N_d=2\cdot\varepsilon^{-2}.
$$  
Notice that we use less values for $\varepsilon$
in case of the standard approach,
since the computing time for $\varepsilon=2^{-5}$
in the standard approach
is already much higher than that
in the regression-based approach
for $\varepsilon=2^{-6}$,
with comparable values of the estimated
root mean squared errors (RMSE)
$\sqrt{\EE\left[|V_{\Delta,N,N_d}-V_{\Delta,0}|^2\right]}$
and
$\sqrt{\EE\left[|\tilde V_{N,N_d,K}-V_{\Delta,0}|^2\right]}$. In addition, we implement the multilevel approach from \cite{BSD} in the following way: set $L=-\log_2(\varepsilon)-2$ for $\varepsilon=2^{-i}$, $i\in\left\{2,3,4,5\right\}$ and choose $(N_d)_l=48\cdot 4^l$ and $N_l=2^{16-l}$ for $l=0,\ldots,L$. Run the multilevel algorithm until the level $L$ is reached. Thus, the cost is of order $\sum_{l=0}^L (N_d)_lN_l=O(2^L)=O(\varepsilon^{-1})$, similar to the one of the regression-based approach. 

Below, we compute the numerical complexities, given 500 independent simulations, and compare it with the theoretical ones, namely, $O(\varepsilon^{-2})$ for the standard approach and $O(\varepsilon^{-1})$ for the multilevel and regression-based approaches
(``limiting formulas'' as $\beta,\kappa\to\infty$). Note that we compute the regression estimates for $v_j(x)$ by means of the algorithm of Tsitsiklis and Van Roy (see \cite{TV}, \cite{TV2}
or Section~8.6 in \cite{Gl}), given $5\cdot 10^4$ independent paths and $\frac{(d+1)(d+2)}{2}+1$ basis functions (polynomials
of $d$ variables up to degree~2 as well as the function~$g_j$) for all the standard, regression-based and multilevel approaches. Further, due to practical purposes, we do not allow to exercise at time $t=0$, which gives us a modified price, namely
$$V_{\Delta,0}=\EE\left[ \max_{1\leq j\leq J}\left(
g_{j}(X_{\Delta,j\Delta})-Y_{\Delta,j\Delta} \right) \right].
$$

\subsection{Two-dimensional example}
We consider the following SDE for $d=m=2$ ($Q=4$)
\begin{align*}
dX_t^i=(r-\delta^i)X_t^idt+\sigma^iX_t^idW_t^i,\quad t\in\left[0,1\right],\quad i=1,2,
\end{align*}
where $r=0$ and $x_0^i=100$, $\sigma^i=0.2$, $\delta^i=0.02$, for $i=1,2$. Hence, $X_t^1$ and $X_t^2$ are two independent geometric Brownian motions.
Further, we consider the Bermudan max-call option
with strike price $100$ and $20$ exercise opportunities
($J=20$), that is,
$g_j\left(x\right)=\max\left\{\max\left\{x_1,x_2\right\}-100,0\right\}$,
$x=(x_1,x_2)$,
for all~$j$.
The ``true'' upper bound $V_{\Delta,0}\approx 12.57$ is estimated as the mean value of \(100\) independent computations of $V_{\Delta,N,N_d}$ with $N=N_d=5\cdot 10^4$.

\begin{figure}[htb!]
\centering
\includegraphics[width=0.7\textwidth]{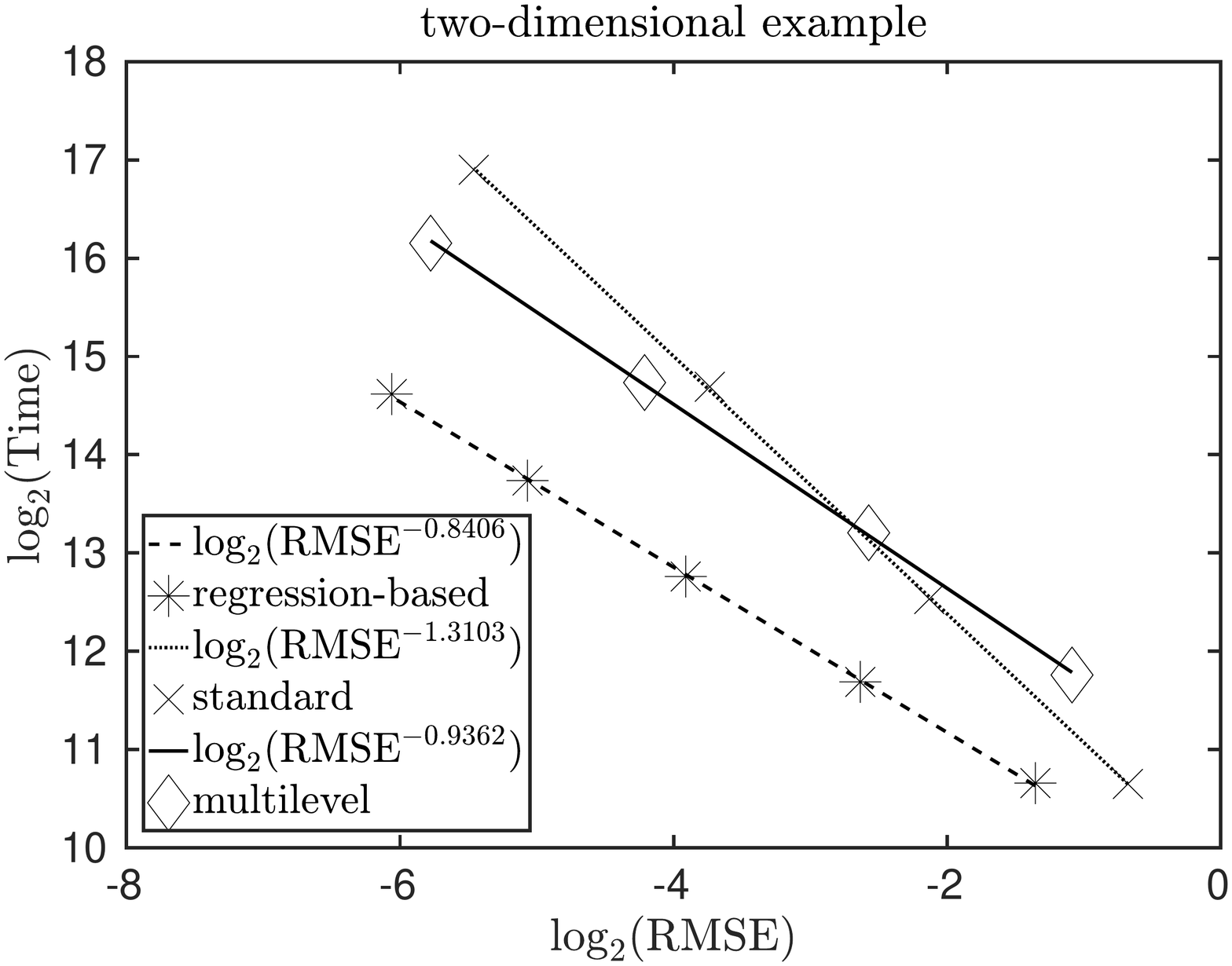}
\includegraphics[width=0.7\textwidth]{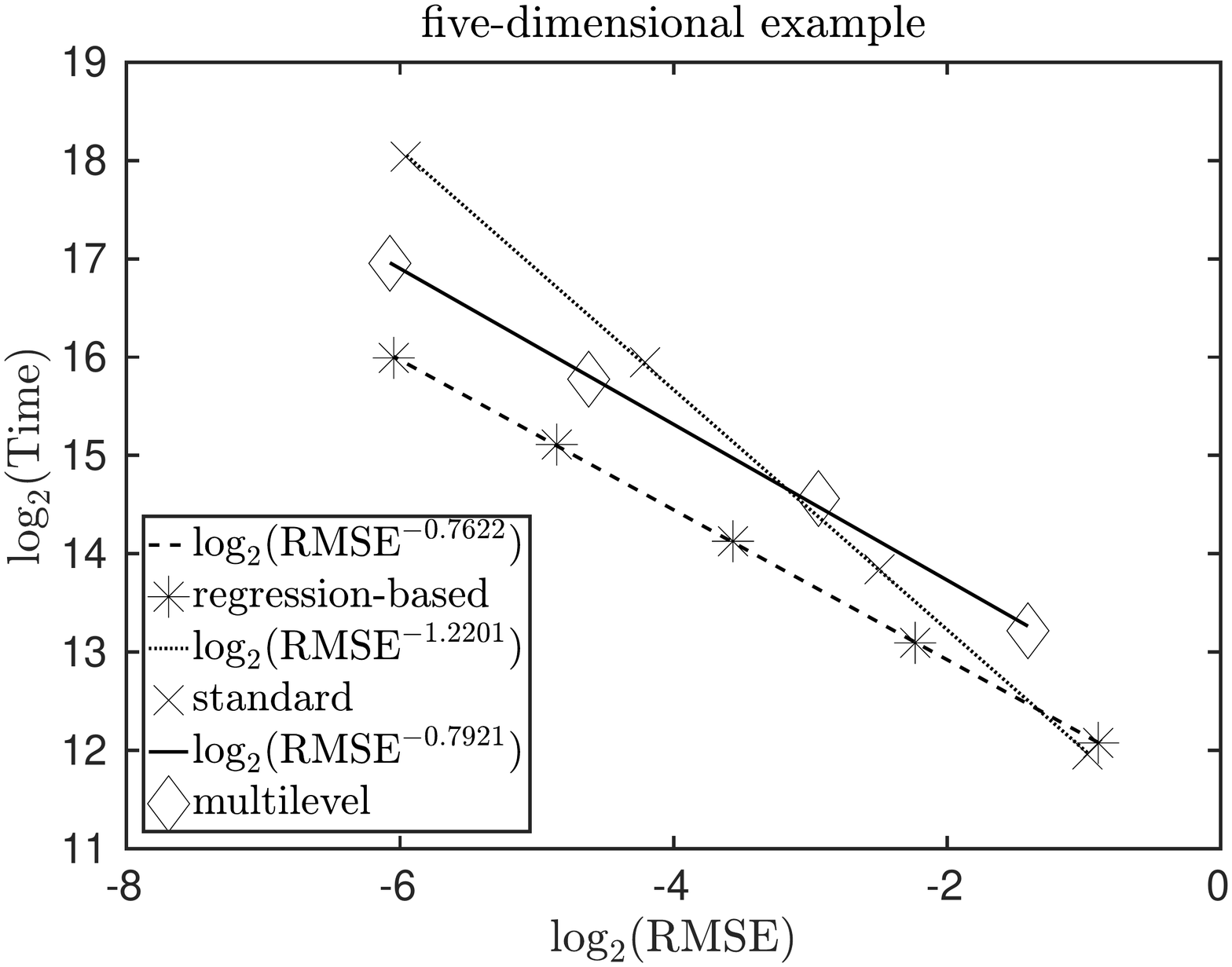}

\caption{Numerical complexities of the regression-based, standard and multilevel approaches in the two- and five-dimensional cases.}
\label{berm:compld}
\end{figure}

As can be seen from the first plot in Figure~\ref{berm:compld},
the estimated numerical complexity is about $\text{RMSE}^{-0.84}$ for the regression-based approach, $\text{RMSE}^{-1.31}$ for the standard approach and $\text{RMSE}^{-0.94}$ for the multilevel approach.
(We speak about numerically estimated RMSEs here.)
The reason for the somewhat unexpected slope $1.31$
in the standard approach is that,
in this numerical example,
the numerical MSE
turned out to be strictly smaller than
the left-hand side of~\eqref{berm:condition_M},
which is of course possible in specific examples.
(Indeed, from the plot corresponding to the standard approach
we get $\text{RMSE}\asymp\varepsilon^{2/1.31}$,
that is, $\text{MSE}\asymp\varepsilon^{4/1.31}$,
which is smaller than $\const/N_d\asymp\varepsilon^2$.)
We see that the regression-based approach works nicely, 
and we can save much computing time as compared to the standard and multilevel approaches to obtain similar accuracies.

\subsection{Five-dimensional example}
We consider the following SDE for $d=m=5$ ($Q=7$)
\begin{align*}
dX_t^i=(r-\delta^i)X_t^idt+\sigma^iX_t^iA^idW_t,\quad t\in\left[0,1\right],\quad i=1,\ldots,5,
\end{align*}
where $r=0$, $x_0^i=100$, $\sigma^i=0.2$, $\delta^i=0.02$ $\forall i$, and $A^i:=\begin{pmatrix}
A^{i,1} \cdots A^{i,5}\end{pmatrix}$, $AA^T=\left(\rho_{ik}\right)_{i,k=1,\ldots,5}$ with $\rho_{ik}=\rho_{ki}\in\left[-1,1\right]$ and $\rho_{ik}=1$ for $i=k$ (that is,
$A^i W$, $i=1,\ldots,5$, are correlated Brownian motions).
For $i<k$ we choose 
\begin{align*}
\rho_{ik}=\left\{\begin{array}{lllrll}
0.9 & \text{if }i=1,\,k=2, && -0.5 & \text{if }i=3,\,k=4,\\
0.2 & \text{if }i\in\left\{1,2,3\right\},\,k=5, &&
-0.2 & \text{if }i=4,\,k=5,\\
0 & \text{otherwise}.\end{array}\right.
\end{align*}
Again, we consider the Bermudan max-call option with strike price $100$, but with only $10$ exercise opportunities ($J=10$), that is,
$g_j\left(x\right)=\max\left\{\max_{i\in\left\{1,\ldots,5\right\}}x_i-100,0\right\}$, for all~$j$, and estimate the upper bound $V_{\Delta,0}\approx 21.07$
via \(100\) independent simulations of $V_{\Delta,N,N_d}$
with $N=N_d=5\cdot 10^4$.

Our empirical findings are illustrated in the second plot in Figure~\ref{berm:compld}.
We observe the numerical complexities of order $\text{RMSE}^{-0.76}$ for the regression-based approach, $\text{RMSE}^{-1.22}$ for the standard approach and $\text{RMSE}^{-0.79}$ for the multilevel approach. Even though the numerical complexities of the regression-based and multilevel approaches are close to each other, we observe that the computing time in case of the regression-based approach is much smaller than the multilevel one, whereas the RMSEs are in a similar region.
As in the previous example,
the regression-based approach shows
a significant complexity reduction effect and outperforms the standard and multilevel approaches numerically.

\section{Proofs}
\subsection*{Proof of Theorem~\protect\ref{berm:prop_vnm}}
In what follows, conditioning on $X^{(n)}_\cdot$
is a shorthand for conditioning on
$\sigma(X^{(n)}_j,0\le j\le J)$.
We set
$$
Y^{(n)}_j:=\EE\left[Y_{j,n,N_d} | X^{(n)}_\cdot\right]
$$
and observe that
$$
Y^{(n)}_j=\sum_{l=1}^j
\left(v_l(X^{(n)}_l)
-\EE[v_l(X^{(n)}_l) | X^{(n)}_{l-1}]\right),
\quad j=0,\ldots,J,
$$
in particular, the process $(Y^{(n)}_j)$
has the same distribution as $(Y_j)$.
Further, we have
\begin{align*}
\EE\left[V_{N,N_d} | X^{(n)}_\cdot\right]
&=\frac1N \sum_{n=1}^N
\EE\left[\maxj\left(
g_j(X^{(n)}_j)-Y_{j,n,N_d}
\right) | X^{(n)}_\cdot\right]\\
&\ge\frac1N \sum_{n=1}^N
\maxj\EE\left[
g_j(X^{(n)}_j)-Y_{j,n,N_d} | X^{(n)}_\cdot\right]\\
&=\frac1N \sum_{n=1}^N
\maxj\left(
g_j(X^{(n)}_j)-Y^{(n)}_j\right),
\end{align*}
which implies the required inequality
$\EE V_{N,N_d}\ge V_0$
by taking expectations of both sides.

For each $n\in\{1,\ldots,N\}$,
we now introduce the filtration
$(\ol\cF^{(n)}_j)_{j\in\{0,\ldots,J\}}$ via
$\ol\cF^{(n)}_0=\mathrm{triv}$ and
$\ol\cF^{(n)}_j=\sigma(X^{(n)}_1,\ldots,X^{(n)}_j,
X_1^{(n_d,n)},\ldots,X_j^{(n_d,n)},n_d=1,\ldots,N_d)$,
$j\in\{1,\ldots,J\}$.
Next, we have
\begin{align}
\label{berm:mse}
\EE\left[(V_{N,N_d}-V_0)^2\right]&=\left(\EE V_{N,N_d}-V_0\right)^2+\Var\left[V_{N,N_d}\right]\\
\notag
&=\left(\EE\left[V_{N,N_d}-\frac1N\sum_{n=1}^N
\maxj\left(g_j(X^{(n)}_j)-Y^{(n)}_j\right)\right]\right)^2+\Var\left[V_{N,N_d}\right].
\end{align}
For the first term in~\eqref{berm:mse}, that is,
for the squared bias, we obtain
\begin{align*}
&\left(\EE V_{N,N_d}-V_0\right)^2\\
\le&\EE\left[\left(V_{N,N_d}-\frac1N\sum_{n=1}^N
\maxj\left(g_j(X^{(n)}_j)-Y^{(n)}_j\right)\right)^2\right]\\
\le &\frac1N\sum_{n=1}^N
\EE\left[\left(
\maxj\left(g_j(X^{(n)}_j)-Y_{j,n,N_d}\right)
-\maxj\left(g_j(X^{(n)}_j)-Y^{(n)}_j\right)
\right)^2\right]\\
\le &\frac1N\sum_{n=1}^N
\EE\maxj\left[\left(
Y_{j,n,N_d}-Y^{(n)}_j
\right)^2\right],
\end{align*}
where we used
$(\frac1N\sum_{n=1}^N a_n)^2\le\frac1N\sum_{n=1}^N a_n^2$
in the first inequality and
$$|\max_j a_j-\max_j b_j|\le\max_j |a_j-b_j|$$
in the second one.
Since $(Y_{j,n,N_d}-Y^{(n)}_j)$
is an $(\ol\cF^{(n)}_j)$-martingale,
Doob's $L^2$ inequality yields
$$\EE\maxj [(Y_{j,n,N_d}-Y^{(n)}_j)^2]
\le4\EE[(Y_{J,n,N_d}-Y^{(n)}_J)^2],$$
so that we get
$$
\left(\EE V_{N,N_d}-V_0\right)^2\le\frac4N\sum_{n=1}^N
\EE\left[\left(
Y_{J,n,N_d}-Y^{(n)}_J
\right)^2\right].
$$
Proceeding as follows
\begin{align*}
\EE\left[\left(Y_{J,n,N_d}-Y^{(n)}_J\right)^2\right]
&=\EE\left[\Var\left[Y_{J,n,N_d} | X^{(n)}_\cdot\right]\right]\\
&=\EE\left[\Var\left[\left.
\sum_{l=1}^J \frac{1}{N_d} \sum_{n_d=1}^{N_d}
v_l(X^{(n_d,n)}_l) \right| X^{(n)}_\cdot
\right]\right]\\
&=\EE\left[\sum_{l=1}^J \frac{1}{N_d} \Var\left[
v_l(X^{(n)}_l) | X^{(n)}_{l-1}
\right]\right]\\
&=\frac{1}{N_d} \sum_{l=1}^J
\EE\left[\Var\left[v_l(X_l) | X_{l-1}\right]\right],
\end{align*}
we obtain the upper bound for the squared bias
\begin{equation}\label{berm:eq:07092016a3}
\left(\EE V_{N,N_d}-V_0\right)^2\le\frac{4}{N_d}\sum_{l=1}^J
\EE\left[\Var\left[v_l(X_l) | X_{l-1}\right]\right].
\end{equation}

Recall the almost sure property of the Doob martingale
$$
Y^{*,(1)}_j:=\sum_{l=1}^{j}\left( v^*_{l}(X_{l}^{(1)})-\EE\left[ v^*_{l}(X_{l}^{(1)})|
X_{l-1}^{(1)}\right] \right),
$$
which here takes the form
$$
\max_{0\le j\le J}
\left(g_j(X_j^{(1)})-Y_j^{*,(1)}\right)=v_0^*(X_0^{(1)})=v^*_0(x_0)
$$
(see~\cite{SchZhH2013}), and, in particular, implies that
the random variable
$\max_{0\le j\le J}(g_j(X_j^{(1)})-Y_j^{*,(1)})$
is, in fact, deterministic.
We, therefore, derive
\begin{equation}\label{eq:17012018a1}
\begin{split}
\Var\left[V_{N,N_d}\right]&=\frac1N\Var
\left[\maxj\left(g_j(X_j^{(1)})-Y_{j,1,N_d}\right)\right]\\
&=\frac1N\Var
\left[\maxj\left(g_j(X_j^{(1)})-Y_{j,1,N_d}\right)
-\maxj\left(g_j(X_j^{(1)})-Y^{*,(1)}_j\right)\right]\\
&\le\frac1N\EE\maxj
\left[\left(Y^{*,(1)}_j-Y_{j,1,N_d}\right)^2\right],
\end{split}
\end{equation}
for the second term in~\eqref{berm:mse}.
Again using Doob's $L^2$ inequality
together with the fact that
martingale differences are uncorrelated, we get
\begin{align*}
&\Var\left[V_{N,N_d}\right]\le\frac4N\EE\left[\left(
Y^{*,(1)}_J-Y_{J,1,N_d}\right)^2\right]
=\frac4N \Var\left[Y^{*,(1)}_J-Y_{J,1,N_d}\right]
\\
&=\frac4N\sum_{l=1}^J
\Var\left[v^*_l(X_l^{(1)})-v_l(X_l^{(1)})
-\EE\left[v^*_l(X_l^{(1)})| X_{l-1}^{(1)}\right]+\frac{1}{N_d}\sum_{n_d=1}^{N_d} v_{l}(X^{(n_d,1)}_{l})
\right]\\
&=\frac4N\sum_{l=1}^J
\EE\left[\Var\left[\left.v^*_l(X_l^{(1)})-v_l(X_l^{(1)})+\frac{1}{N_d}\sum_{n_d=1}^{N_d} v_{l}(X^{(n_d,1)}_{l}) \right| X_{l-1}^{(1)}\right]\right].\\
\end{align*}
Since, conditionally on $X^{(1)}_{l-1}$,
the random variables
$X^{(1)}_l, X^{(1,1)}_l,\ldots,X^{(N_d,1)}_l$
are independent, we arrive at
\begin{align*}
&\Var\left[V_{N,N_d}\right]\\
&\le\frac4N\sum_{l=1}^J\left(
\EE\left[\Var\left[\left.v^*_l(X_l^{(1)})-v_l(X_l^{(1)}) \right| X_{l-1}^{(1)}\right]\right]+\EE\left[\Var\left[\left.\frac{1}{N_d}\sum_{n_d=1}^{N_d} v_{l}(X^{(n_d,1)}_{l}) \right| X_{l-1}^{(1)}\right]\right]\right)\\
&=\frac4N\sum_{l=1}^J\left(
\EE\left[\Var\left[v^*_l(X_l)-v_l(X_l) | X_{l-1}\right]\right]+\frac{1}{N_d}\EE\left[\Var\left[v_{l}(X_{l}) | X_{l-1}\right]\right]\right)\\
&=\frac1N\frac4{N_d}\sum_{l=1}^J \EE\left[\Var\left[v_{l}(X_{l}) | X_{l-1}\right]\right]
+\frac4N\sum_{l=1}^J \EE\left[\Var\left[v^*_l(X_l)-v_l(X_l) | X_{l-1}\right]\right].
\end{align*}
Together with \eqref{berm:mse}
and~\eqref{berm:eq:07092016a3},
we obtain first inequality in~\eqref{berm:eq:bound_vnm}.
The second one now follows from
$$
\EE\left[\Var\left[v^*_l(X_l)-v_l(X_l) | X_{l-1}\right]\right]
\le\EE\left[\left(v^*_l(X_l)-v_l(X_l)\right)^2\right].
$$

\subsection*{Proof of Proposition~\protect\ref{prop:04032018a1}}
We set
$$
Z_j^{(n)}:=\EE\left[Z_{j,n,N_d}|X^{(n)}_\cdot\right]
$$
and observe that
$$
Z_j^{(n)}=\EE\left[v_j(X_j^{(n)})|X_{j-1}^{(n)}\right],
\quad j=1,\ldots,J.
$$
We also define
$$
U_N:=\frac1N\sum_{n=1}^N
\left[g_J(X_J^{(n)})+\sum_{j=1}^J
\left(g_{j-1}(X^{(n)}_{j-1})-Z_j^{(n)}\right)^+\right]
$$
and notice that $\EE U_N=U_0$.
By Jensen's inequality
$$
\EE\left[U_{N,N_d}|X^{(n)}_\cdot\right]\ge U_N,
$$
which, in turn, implies~\eqref{eq:04032018a1}.

Next, we apply the formula
$$
\EE\left[(U_{N,N_d}-U_0)^2\right]=\left(\EE U_{N,N_d}-U_0\right)^2+\Var\left[U_{N,N_d}\right]
$$
and notice that the second term here
is precisely the second term
on the right-hand side of~\eqref{eq:04032018a2}.
It remains to prove that
$(\EE U_{N,N_d}-U_0)^2$
is equal to or less than the first term
on the right-hand side of~\eqref{eq:04032018a2}.
To this end, we sketch the main steps as follows:
\begin{equation*}
\begin{split}
\left(\EE U_{N,N_d}-U_0\right)^2
&\le\EE\left[(U_{N,N_d}-U_N)^2\right]\\
&\le\frac1N\sum_{n=1}^N
\EE\left[\bigg(\sum_{j=1}^J\left[
\left(g_{j-1}(X^{(n)}_{j-1})-Z_{j,n,N_d}\right)^+
- \left(g_{j-1}(X^{(n)}_{j-1})-Z_j^{(n)}\right)^+
\right]\bigg)^2\right]\\
&\le J\sum_{j=1}^J
\EE\left[\left(Z_{j,1,N_d}-Z_j^{(1)}\right)^2\right]\\
&=J\sum_{j=1}^J
\EE\left[\bigg(
\frac1{N_d}\sum_{n_d=1}^{N_d}
v_j(X_j^{(n_d,1)})-Z_j^{(1)}
\bigg)^2\right]\\
&=J\sum_{j=1}^J
\EE\left[\Var\left[\left.
\frac1{N_d}\sum_{n_d=1}^{N_d} v_j(X_j^{(n_d,1)})
\right|X_{j-1}^{(1)}\right]\right]\\
&=\frac J{N_d}\sum_{j=1}^J
\EE\left[\Var\left[v_{j}(X_{j}) | X_{j-1}\right]\right].
\end{split}
\end{equation*}
This completes the proof.

\subsection*{Proof of Theorem~\protect\ref{berm:cor: expansion_main}}
The expansion obviously holds for $p=1$ and $j=0$.
Indeed, due to the orthonormality and completeness
of the system $\left(\phi_{k}\right)$, we have 
\[
f(X_{\Delta,\Delta})=\mathrm{\EE}\left[f(X_{\Delta,\Delta})\right]+\sum_{k\geq1}a_{1,1,k}(x_0)\phi_{k}(\xi_{1})
\]
with 
\[
a_{1,1,k}(x_0)=\mathrm{\EE}\left[f(X_{\Delta,\Delta})\phi_{k}\left(\xi_{1}\right)\right],
\]
provided $\EE\left[\left|f(X_{\Delta,\Delta})\right|^{2}\right]<\infty.$ 
Recall that $\cG_l=\sigma(\xi_{1},\ldots,\xi_{l}),$
$l=1,2,\ldots,L$,
and $\cG_0=\mathrm{triv}$.
Suppose that (\ref{berm:eq:expansioN_rond})
holds for $p=q$, all $j<q$, and all Borel-measurable functions
$f$ with $\EE\left[|f(X_{\Delta,q\Delta})|^2\right]<\infty$.
Let us prove it for $p=q+1$.
Given $f$ with $\EE\left[|f(X_{\Delta,p\Delta})|^2\right]<\infty$,
due to the orthonormality and completeness
of the system $\left(\phi_{k}\right)$, we get by
conditioning on $\mathcal{G}_{q}$,
$$
f(X_{\Delta,p\Delta})=\mathrm{\EE}\left[\left.f(X_{\Delta,p\Delta})\right|\cG_q\right]+\sum_{k\geq1}\alpha_{p,q+1,k}\phi_{k}(\xi_{q+1}),
$$
where 
\begin{equation*}
\alpha_{p,q+1,k}
=\mathrm{\EE}\left[\left.f(X_{\Delta,p\Delta})\phi_{k}(\xi_{q+1})\right|\mathcal G_q\right].
\end{equation*}
By the Markov property of $(X_{\Delta,l\Delta})$,
we have
$\EE[f(X_{\Delta,p\Delta})|\cG_q]
=\EE[f(X_{\Delta,p\Delta})|X_{\Delta,q\Delta}]$.
Furthermore, a calculation involving 
intermediate conditioning on $\mathcal G_{q+1}$
and the recurrence relation
$X_{\Delta,(q+1)\Delta}=\Phi_{q+1}(X_{\Delta,q\Delta},\xi_{q+1})$
verifies that
$$
\alpha_{p,q+1,k}=\mathrm{\EE}\left[\left.f(X_{\Delta,p\Delta})\phi_{k}(\xi_{q+1})\right|X_{\Delta,q\Delta}\right]
=a_{p,q+1,k}(X_{\Delta,q\Delta})
$$
for suitably chosen
Borel-measurable functions $a_{p,q+1,k}$.
We thus arrive at
\begin{align}
\label{berm:sig_X}
f(X_{\Delta,p\Delta})=\mathrm{\EE}\left[\left.f(X_{\Delta,p\Delta})\right|X_{\Delta,q\Delta}\right]+\sum_{k\geq1}a_{p,q+1,k}(X_{\Delta,q\Delta})\phi_{k}(\xi_{q+1}),
\end{align}
which is the required statement in the case $j=q$.
Now assume $j<q$.
The random variable
$\mathrm{\EE}\left[\left.f(X_{\Delta,p\Delta})\right|X_{\Delta,q\Delta}\right]$
is square integrable and has the form
$g(X_{\Delta,q\Delta})$,
hence the induction hypothesis applies, and we get
\begin{equation}\label{eq:28082017a1}
\mathrm{\EE}\left[\left.f(X_{\Delta,p\Delta})\right|X_{\Delta,q\Delta}\right]=\mathrm{\EE}\left[\left.f(X_{\Delta,p\Delta})\right|X_{\Delta,j\Delta}\right]+\sum_{k\geq1}\sum_{l=j+1}^{q}a_{p,l,k}(X_{\Delta,(l-1)\Delta})\phi_{k}(\xi_{l})
\end{equation}
with 
\begin{align*}
a_{p,l,k}(X_{\Delta,(l-1)\Delta}) &= \EE\left[\left.\EE\left[\left.f(X_{\Delta,p\Delta})\right|\mathcal{G}_{q}\right]\phi_{k}(\xi_{l})\right|\mathcal{G}_{l-1}\right]
= \EE\left[\left.f(X_{\Delta,p\Delta})\phi_{k}(\xi_{l})\right|\mathcal{G}_{l-1}\right]\\
&=\EE\left[\left.f(X_{\Delta,p\Delta})\phi_{k}(\xi_{l})\right|X_{\Delta,(l-1)\Delta}\right].
\end{align*}
Formulas \eqref{berm:sig_X}
and~\eqref{eq:28082017a1} conclude the proof.

\subsection*{Proof of Theorem~\protect\ref{berm:theorem:regression_cv}}
It holds 
\begin{equation}\label{eq:29082017a2}
\EE\left[\Var\left[\left.
f(X_{\Delta,p\Delta})-\tilde M_{j,p,K}
\right|X_{\Delta,j\Delta}\right]\right]
=\EE\left[\left|M_{j,p}-\tilde{M}_{j,p,K}\right|^{2}\right].
\end{equation}
We have
\begin{equation}
\begin{split}
\EE\left[\left|M_{j,p}-\tilde{M}_{j,p,K}\right|^{2}\right] & =\EE\left[\left|\sum_{k=K+1}^{\infty}\sum_{l=j+1}^{p}a_{p,l,k}(X_{\Delta,(l-1)\Delta})\phi_{k}(\xi_{l})\right|^{2}\right]\\
 &\phantom{=} +\EE\left[\left|\sum_{k=1}^{K}\sum_{l=j+1}^{p}\left(a_{p,l,k}(X_{\Delta,(l-1)\Delta})-\tilde a_{p,l,k}(X_{\Delta,(l-1)\Delta})\right)\phi_{k}(\xi_{l})\right|^{2}\right]\\
 & =\sum_{k=K+1}^{\infty}\sum_{l=j+1}^{p}\EE\left[a_{p,l,k}^{2}(X_{\Delta,(l-1)\Delta})\right]\\
 &\phantom{=} +\sum_{k=1}^{K}\sum_{l=j+1}^{p}\EE\left[\left(a_{p,l,k}(X_{\Delta,(l-1)\Delta})-\tilde a_{p,l,k}(X_{\Delta,(l-1)\Delta})\right)^{2}\right].
\end{split}
\end{equation}
It follows from Theorem~11.3 in \cite{gyorfi2002distribution} that
\begin{equation}\label{eq:29082017a3}
\begin{split}
&\EE\left[\left(a_{p,l,k}(X_{\Delta,(l-1)\Delta})-\tilde a_{p,l,k}(X_{\Delta,(l-1)\Delta})\right)^{2}\right]  \\
&\leq\tilde c F^2\frac{Q(\log(N_r)+1)}{N_r}+8\inf_{\psi\in\mathrm{span}(\psi_{1},\ldots,\psi_{Q})}\EE\left[\left|a_{p,l,k}(X_{\Delta,(l-1)\Delta})-\psi(X_{\Delta,(l-1)\Delta})\right|\right]^{2}
\end{split}
\end{equation}
for some universal constant $\tilde c$, since 
\[
\Var\left[\left.f(X_{\Delta,p\Delta})\phi_{k}\left(\xi_{l}\right)\right|X_{\Delta,(l-1)\Delta}=x\right]\leq F^{2}
\]
and
\begin{align*}
\left|\EE\left[\left.f(X_{\Delta,p\Delta})\phi_{k}\left(\xi_{l}\right)\right|X_{\Delta,(l-1)\Delta}=x\right]\right|\leq F.
\end{align*}
The result now follows from
\eqref{berm:ass}--\eqref{eq:29082017a4} and
\eqref{eq:29082017a2}--\eqref{eq:29082017a3}.

\subsection*{Proof of Theorem~\protect\ref{berm:prop:29092016a1}}
By the same calculation as the one leading
to~\eqref{berm:eq:bound_vnm}
(see the proof of Theorem~\ref{berm:prop_vnm}),
we get
\begin{align*}
\EE\left[(\tilde V_{N,N_d,K}-V_{\Delta,0})^2\right]
&\leq\frac{4}{N_d}\left(1+\frac{1}{N}\right)
\sum_{l=1}^J
\EE\left[\Var\left[
v_l(X_{\Delta,l\Delta})-\tilde M_{l,K} | X_{\Delta,(l-1)\Delta}
\right]\right]\\
&\phantom{=}+\frac4N
\sum_{l=1}^{J}\EE\left[\left(v_{l}^{*}(X_{\Delta,l\Delta})-v_{l}(X_{\Delta,l\Delta})\right)^{2}\right].
\end{align*}
It remains to apply Theorem~\ref{berm:theorem:regression_cv}
to the first term on the right-hand side.

\section{Appendix: derivation of complexity~\eqref{berm:compl_sol}}
\label{sec:append_compl}
Let us, for simplicity, first ignore the $\log(N_r)$-term in~\eqref{berm:mse_constr} and consider only the terms
w.r.t. the variables $K,Q,N,N_d,N_r$ which shall be optimized, since the constants $J,B_\beta,D_\kappa$ do not affect the terms on $\varepsilon$. Further, we consider the log-cost and log-constraints 
rather than~\eqref{berm:cost_algorithm} and~\eqref{berm:mse_constr}. Let us subdivide the optimization problem into three cases:

\begin{enumerate}
\item[a)] $\max\left\{N_rQ^2,NQ,NN_d\right\}=N_rQ^2$.
Here, we have the Lagrange function
\begin{align}
\notag
L(K,Q,N,N_d,N_r):=&\log(K) + \log(N_r) + 2\log(Q)+ \lambda_1(\log(N)-\log(N_r)- \log(Q)) \\
\notag
\phantom{:=}&+ \lambda_2(\log(N) + \log(N_d) - \log(N_r) – 2\log(Q))\\
\notag
\phantom{:=}&+ \lambda_3(\log(K)+\log(Q) - \log(N_d) - \log(N_r) - 2\log(\varepsilon))  \\
\notag
\phantom{:=}&+ \lambda_4(-\log(N_d) - \beta\log(K)- 2\log(\varepsilon) ) \\
\notag
\phantom{:=}&+ \lambda_5(\log(K) - \log(N_d)  - \kappa\log(Q)- 2\log(\varepsilon)) \\
\phantom{:=}&
+ \lambda_6(-\log(N) - 2\log(\varepsilon)).
\end{align}
Considering $\frac{\partial L}{\partial K}=\frac{\partial L}{\partial Q}=\frac{\partial L}{\partial N}=\frac{\partial L}{\partial N_d}=\frac{\partial L}{\partial N_r}=0$ leads to the unique solution $\lambda_1=0,\lambda_i>0$ for $i>1$. More precisely, due to five equations with six variables $\lambda_1,\ldots,\lambda_6$, at least one $\lambda_i$ has to be zero. In the others cases $\lambda_i=0$ for $i>1$, we either obtain $\lambda_k<0$ for some $k\neq i$, or the  corresponding constraint to $\lambda_i$ is not satisfied (that is $>0$) such that these solutions are not optimal, respectively they do not satisfy all constraints.
Since we have $\lambda_i>0$ for all $i>1$, all constraints corresponding to those $\lambda_i$ have to be active
(that is, zero). This gives us
\begin{align*}
K=O\left(\varepsilon^{-\frac{4\kappa}{3\beta + \kappa + 2\beta\kappa + 3}}\right),\quad Q=O\left(\varepsilon^{-\frac{4(1+\beta)}{3\beta + \kappa + 2\beta\kappa + 3}}\right),\quad N=O\left(\varepsilon^{-2}\right),\\ N_d=O\left(\varepsilon^{-\frac{6(1+\beta)+2\kappa}{3\beta + \kappa + 2\beta\kappa + 3}}\right),\quad N_r=O\left(\varepsilon^{-\frac{4(1+\beta)(1+\kappa)}{3\beta + \kappa + 2\beta\kappa + 3}}\right),
\end{align*}
provided that\footnote{The condition $\beta>1$ is required to satisfy $\lambda_i>0$ for all $i>1$.}~$\beta>1$. Hence, the complexity is 
\begin{align}
\label{complexity_deriv_a}
\mathcal{C}=O\left(KN_rQ^2\right)=O\left(\varepsilon^{-\frac{4(1+\beta)(3+\kappa)+4\kappa}{3\beta + \kappa + 2\beta\kappa + 3}}\right).
\end{align}

\item[b)]
$\max\left\{N_rQ^2,NQ,NN_d\right\}=NQ$. Here, we obtain, similarly to case~a), the complexity of order
$$
O\left(\varepsilon^{-2-\frac{4 (\beta + \kappa + 1)}{\beta (\kappa + 1) + 1}}\right)
$$
for $\beta>1$, which is worse than~\eqref{complexity_deriv_a}.

\item[c)]
$\max\left\{N_rQ^2,NQ,NN_d\right\}=NN_d$. This gives us the same result as in case~a).
Hence, the complexity in~\eqref{complexity_deriv_a}
is the overall optimal solution.
\end{enumerate}

In the case $\beta=1$, we get (compare with~\eqref{berm:mse_constr})
$$
N\gtrsim \varepsilon^{-2},\quad KN_d\gtrsim\varepsilon^{-2},
$$
and thus 
$$\mathcal{C}\gtrsim KNN_d\gtrsim\varepsilon^{-4},$$
which means that the complexity is always worse than the one of the standard nested simulations approach
discussed in Section~\ref{berm:sec:ns}.
Next we consider also the remaining terms $J,B_\beta,D_\kappa$ and arrive at~\eqref{berm:compl_sol} via equalizing all constraints in~\eqref{berm:mse_constr} as well as considering $NN_d=N_rQ^2$ (provided that $\beta>1$). Finally, we add the log-term concerning $\varepsilon$ in the
parameters $N_r$ and $N_d$ to ensure that all constraints are really satisfied. 
Notice that the constant $C_{J,\beta,\kappa}$ in~\eqref{berm:compl_sol} is an upper bound of the constant which arises from equalizing the above mentioned constraints.

\bibliographystyle{abbrv}
\bibliography{dual_lit}

\begin{thebibliography}{10}

\bibitem{AB}
L.~Andersen and M.~Broadie.
\newblock Primal-dual simulation algorithm for pricing multidimensional
  {A}merican options.
\newblock {\em Management Science}, 50(9):1222--1234, 2004.

\bibitem{belomestny2009true}
D.~Belomestny, C.~Bender, and J.~Schoenmakers.
\newblock True upper bounds for {B}ermudan products via non-nested {M}onte
  {C}arlo.
\newblock {\em Mathematical Finance}, 19(1):53--71, 2009.

\bibitem{BHNU}
D.~Belomestny, S.~H\"afner, T.~Nagapetyan, and M.~Urusov.
\newblock Variance reduction for discretised diffusions via regression.
\newblock {\em Preprint, arXiv:1510.03141v4}, 2017.

\bibitem{BMS:2009}
D.~Belomestny, G.~Milstein, and V.~Spokoiny.
\newblock Regression methods in pricing {A}merican and {B}ermudan options using
  consumption processes.
\newblock {\em Quant. Finance}, 9(3):315--327, 2009.

\bibitem{belomestny2006monte}
D.~Belomestny and G.~N. Milstein.
\newblock Monte {C}arlo evaluation of {A}merican options using consumption
  processes.
\newblock {\em International Journal of Theoretical and Applied Finance},
  9(04):455--481, 2006.

\bibitem{BSD}
D.~Belomestny, J.~Schoenmakers, and F.~Dickmann.
\newblock Multilevel dual approach for pricing {A}merican style derivatives.
\newblock {\em Finance and Stochastics}, 17(4):717--742, 2013.

\bibitem{chen2007additive}
N.~Chen and P.~Glasserman.
\newblock Additive and multiplicative duals for {A}merican option pricing.
\newblock {\em Finance and Stochastics}, 11(2):153--179, 2007.

\bibitem{CLP2002}
E.~Cl{\'e}ment, D.~Lamberton, and P.~Protter.
\newblock An analysis of a least squares regression method for {A}merican
  option pricing.
\newblock {\em Finance Stoch.}, 6(4):449--471, 2002.

\bibitem{dupuis2005convergence}
P.~Dupuis and H.~Wang.
\newblock On the convergence from discrete to continuous time in an optimal
  stopping problem.
\newblock {\em The Annals of Applied Probability}, 15(2):1339--1366, 2005.

\bibitem{Gl}
P.~Glasserman.
\newblock {\em {M}onte {C}arlo methods in financial engineering}, volume~53.
\newblock Springer Science \& Business Media, 2003.

\bibitem{gyorfi2002distribution}
L.~Gy{\"o}rfi, M.~Kohler, A.~Krzy{\.z}ak, and H.~Walk.
\newblock {\em A distribution-free theory of nonparametric regression}.
\newblock Springer Series in Statistics. Springer-Verlag, New York, 2002.

\bibitem{HK}
M.~B. Haugh and L.~Kogan.
\newblock Pricing {A}merican options: a duality approach.
\newblock {\em Oper. Res.}, 52(2):258--270, 2004.

\bibitem{KP}
P.~E. Kloeden and E.~Platen.
\newblock {\em Numerical solution of stochastic differential equations},
  volume~23.
\newblock Springer Science \& Business Media, 1992.

\bibitem{LS}
F.~A. Longstaff and E.~S. Schwartz.
\newblock Valuing {A}merican options by simulation: {A} simple least-squares
  approach.
\newblock {\em Review of Financial studies}, 14(1):113--147, 2001.

\bibitem{MilsteinTretyakov:2004}
G.~N. Milstein and M.~V. Tretyakov.
\newblock {\em Stochastic numerics for mathematical physics}.
\newblock Scientific Computation. Springer-Verlag, Berlin, 2004.

\bibitem{Ro}
L.~C.~G. Rogers.
\newblock Monte {C}arlo valuation of {A}merican options.
\newblock {\em Math. Finance}, 12(3):271--286, 2002.

\bibitem{SchZhH2013}
J.~Schoenmakers, J.~Zhang, and J.~Huang.
\newblock Optimal dual martingales, their analysis, and application to new
  algorithms for {B}ermudan products.
\newblock {\em SIAM J. Financial Math.}, 4(1):86--116, 2013.

\bibitem{TV}
J.~N. Tsitsiklis and B.~Van~Roy.
\newblock Optimal stopping of {M}arkov processes: {H}ilbert space theory,
  approximation algorithms, and an application to pricing high-dimensional
  financial derivatives.
\newblock {\em IEEE Transactions on Automatic Control}, 44(10):1840--1851,
  1999.

\bibitem{TV2}
J.~N. Tsitsiklis and B.~Van~Roy.
\newblock Regression methods for pricing complex {A}merican-style options.
\newblock {\em IEEE Transactions on Neural Networks}, 12(4):694--703, 2001.

\bibitem{zanger2013quantitative}
D.~Z. Zanger.
\newblock Quantitative error estimates for a least-squares {M}onte {C}arlo
  algorithm for {A}merican option pricing.
\newblock {\em Finance and Stochastics}, 17(3):503--534, 2013.

\end{thebibliography}
\end{document}